\documentclass[preprint,12pt]{elsarticle}

\usepackage{amsmath,amssymb,amsfonts}
\usepackage{url}
\usepackage{tabularx}
\usepackage{makecell} 
\usepackage{ragged2e} 
\usepackage{booktabs} 
\usepackage[figuresright]{rotating}

\begin{document}

\begin{frontmatter}



\title{Denoising, Segmentation and Volumetric Rendering of Optical Coherence Tomography Angiography (OCTA) Image using Deep Learning Techniques: A Review} 


\author[label1]{Kejie Chen} 
\author[label1]{Guanbin Gao}
\author[label2]{Xiaochun Yang}
\author[label1]{Wenbo Wang}
\author[label1]{Jing Na}
\affiliation[label1]{organization={Department of Mechanical and Electronic Engineering, Kunming University of Science and Technology},
            state={Yunnan},
            country={China}}
\affiliation[label2]{organization={Eye Clinics, Yunnan First People's Hospital},
            state={Yunnan},
            country={China}}

\begin{abstract}
Optical coherence tomography angiography (OCTA) is clinically valuable for diagnosing ocular diseases, offering safety and speed advantages over dye-based angiography while retaining the ability to characterize micro-scale pathological features like microaneurysms and ischemia areas in the fundus. However, inherent noise and artifacts from acquisition devices and protocols compromise data homogeneity, diagnostic accuracy, and repeatability. Deep learning (DL) based imaging analysis methods can automatically detect and remove artifacts, suppress systematic noise, and enhance OCTA data quality, serving as powerful tools for segmenting and identifying both normal and pathological structures. This study reviews state-of-the-art research published within the last five years on applying DL to analyze OCTA data, providing a comprehensive introduction to relevant DL techniques, including model architecture design, objective function design, and training strategies. It first introduces and compares the performance of three fundamental DL model types on OCTA \textit{en face} images, which serve as building blocks for larger advanced models. Next, it classifies state-of-the-art studies by application type: (1) denoising and data enhancement, (2) vasculature and pathological feature segmentation, and (3) volumetric rendering and visualization, discussing each application's DL design principles and practical pros and cons. Lastly, it summarizes publicly available OCTA datasets. Overall, this review provides valuable insights for engineers developing novel DL models by leveraging OCTA data characteristics and disease-specific pathology, while assisting technicians and clinicians in selecting appropriate DL models for fundamental research and disease screening.
\end{abstract}




\end{frontmatter}



\section{Introduction}
\label{Introduction}
Optical coherence tomography angiography (OCTA) is an emerging technique that enables imaging retinal capillary plexuses, choroicapillaris and vascularization in the anterior segments \cite{1,3,4}. The development of OCTA has been around twenty years, while the first demonstration of visualizing the vasculature in the human eye was performed in 2006 \cite{1}. Compared with traditional methods of imaging flow, including fluorescein angiography (FA) and Indocyanine green angiography (ICGA), OCTA has the advantage that it can visualize deeper vasculature with depth resolution, does not require administration of exogenous contrast, and is not obscured by dye leakage \cite{5}. Thus, OCTA has become one of the most widely-used techniques in clinics and laboratory research for monitoring ocular conditions and diagnosing ocular diseases, such as glaucoma, diabetic retinopathy and macular degeneration \cite{2, 6,7,8,9,10,11}.

OCTA utilizes the concept of low coherence interferometry of optical coherence tomography (OCT) to acquire cross sectional (B-scan) images. By performing repeated B-scans in the same location and analyzing the signal decorrelation, the motion contrast is detected for determination of flows (Fig. 1) \cite{13,12}. Fig. 1 is the schematic illustration of the OCTA data. The surface of the eyeball is defined as the \textit{en face} plane, and the depth direction is defined as the direction perpendicular to the \textit{en face} plane (Fig. 1a). As shown in Fig. 1b, the \textit{en face} OCT data contains vascular structures at the same tissue depth. A B-scan slice is the cross-sectional view of the fundus tissue. In the \textit{en face} image, a B-scan slice becomes a 2D curve or a line, such as the green arrow shown in Fig. 1b. After performing the decorrelation analysis of the B-scan time series data, blood flow signals in the fundus are extracted. The OCTA data are the flow signals in the retinal and choroidal tissue volume (Fig. 1c). In addition, the flow signals are used to represent the vasculature by ignoring the thin vascular walls. Most clinically-used equipments provide the superimposition of OCT and OCTA data in one image, as shown in Fig. 1d. The vascular structures at different depths (\textcircled{1},\textcircled{2},\textcircled{3}) can be visualized by maximum intensity projection onto the \textit{en face} plane.

\begin{figure}[!t]
\centerline{\includegraphics[width=0.8\columnwidth]{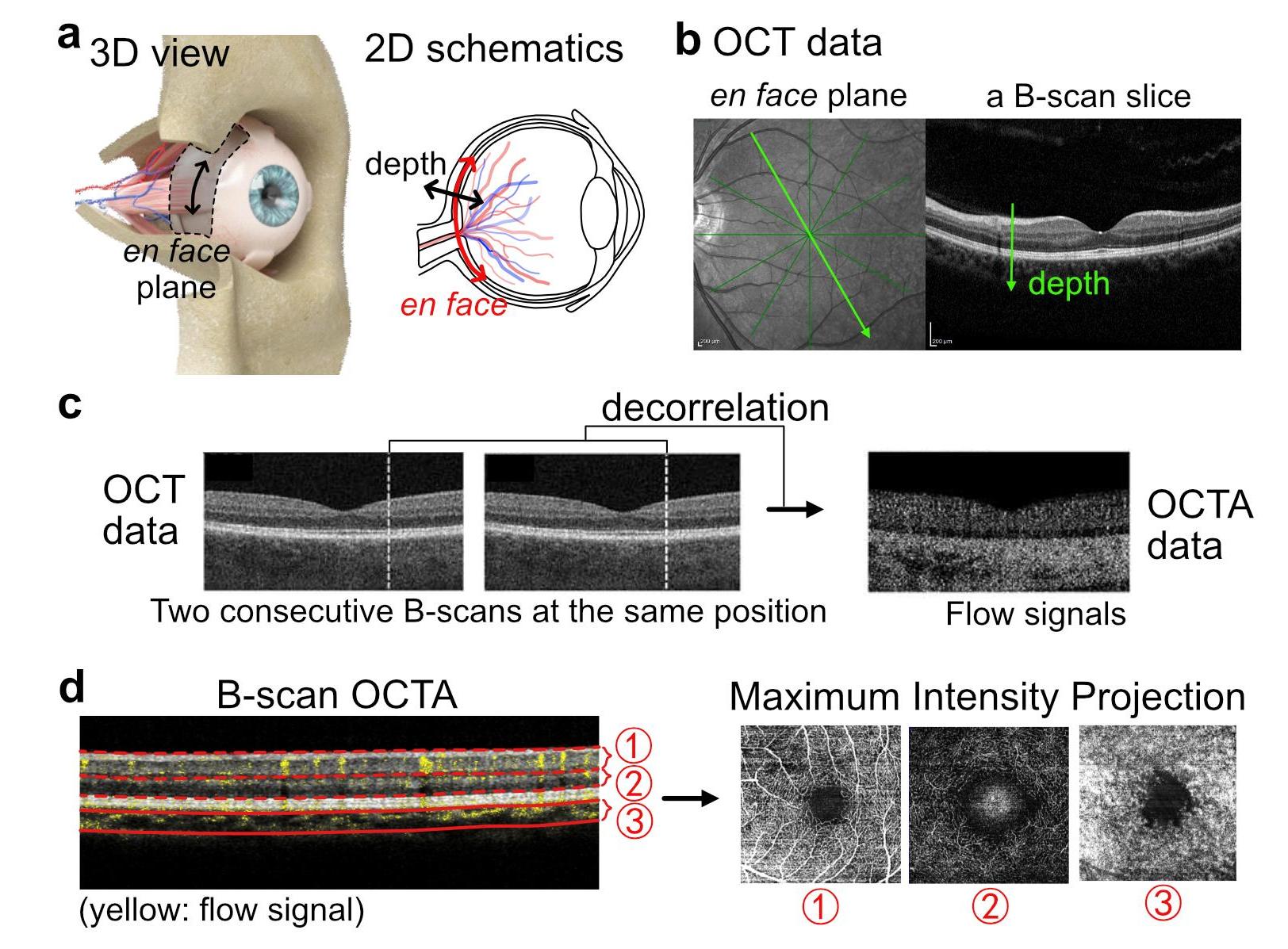}}
\caption{Schematic illustration of the OCTA data. \textbf{(a)} The 3D and 2D view of the \textit{en face} plane and the depth direction. \textbf{(b)} Examples of the \textit{en face} OCT image and the cross-sectional OCT B-scan image. \textbf{(c)} Schematic illustration of measuring the flow signals in one cross-sectional slice by performing the decorrelation of consecutive B-scan OCT signals. \textbf{(d)} Example of the \textit{en face} OCTA images in different tissue depths generated based on the maximum intensity projection of a bulk layer of flow signals.}
\label{fig1}
\end{figure}

Although OCTA has advantages of visualizing the multi-layer micro-vasculatures, the acquired data is highly dependent on the OCTA equipments, scanning protocols and signal processing methods \cite{14,15,16}. There are many more artifacts in the OCTA data compared with FA and ICGA methods, which can easily lead to misinterpretation \cite{1,17,18,19,20,21,22,23,24}. For example, microsaccade or other rapid movements of retina and choroid induced by cardiac cycle and breathing can generate motion artifacts, such as stripes and crisscross of vessels. Because fluctuating motions of retina and choroid are difficult to discern and are often ignored during the signal acquisition process, this kind of motion artifact widely existed in most OCTA data \cite{18,19,20}. Eye blinking, large lipid deposit, cystoid and edema change the refractive index of the tissue, which causes a part of the image to be out of focus or generate variations of signals which are unrelated to the flow \cite{18,19}. Ocular media opacity, edema, hemorrhage, pigment clumping, lesions and other light blockage lead to the low-OCT-signal artifacts \cite{23}. Moreover, when imaging eyes with pathology, the disrupted retinal structures such as the abnormal retinal thickness and the altered retinal curvature make it hard for clinicians and computational algorithms to correctly identify and segment retinal layers \cite{24}. 3D volumetric rendering of the OCTA data is an alternative way to help ophthalmologists visualize the curved retinal tissue layers, cystoid spaces and tumefactions \cite{1,30}.But the 3D reconstruction of vascular plexuses and irregular abnormalities in the fundus is still challenging. Most commercialized OCTA equipments have not implemented with the 3D volumetric rendering function so far \cite{26,27}.

In recent years, deep learning (DL) technologies has led to breakthroughs in the ophthalmology. DL models applied to the fundus photographs have shown great capability and high accuracy of automatically screening and diagnosing vision-threatening diseases, such as diabetic retinopathy (DR), glaucoma and age-related macular degeneration (AMD) \cite{28,30,33}. Color fundus images are relatively easier to acquire in clinics and contain less noises and artifacts \cite{34,35,42}. DL models can be trained by using a large amount of fundus image data, such that the model achieves diagnostic accuracy as good as an experienced human ophthalmologist. OCT data is another data type that can be analyzed effectively by DL models \cite{42,43}. Many studies have demonstrated that the DL models correctly identified retinal layers, inter-tissue fluids and other anatomical and pathological structures in both superficial and deep retinal tissues using the OCT data \cite{37,38,39}. 

Compared with color fundus images and OCT data, OCTA data contains plenty of additional important information in the successive OCT scans. OCTA is particularly useful for studying the dynamical blood flows which provide mass and oxygen supply to the retinal photoreceptor cells \cite{1,3,4}. However, DL models have not been well developed and applied to robustly analyze the OCTA data so far, due to the artifacts and noises, inhomogeneity and class imbalance problems in the data. Several survey papers published recently discussed both the traditional image analysis methods and the artificial intelligence (AI) methods for OCTA data \cite{44,44-1,44-2,44-3,44-4,44-5,44-6}. Three survey papers \cite{44,44-2,44-4} summarized the applications of DL methods including segmentation of vasculature and pathological areas, classification of eye diseases and prediction of disease progression using OCTA data. But how to develop and apply these models were not specifically discussed. Meiburger et al. \cite{44-1} presented detailed discussions of the image analysis algorithms of OCTA data, most of which are traditional statistical methods such as thresholding, edge detection, support vector machine, k-nearest neighbor, etc. Several canonical DL models, including UNet, ResNet, DenseNet and VGG, for OCTA image segmentation and classification are presented as well. Le et al. \cite{44-3} reviewed the quantitative image features and different image modalities that can be utilized in the machine learning models. Furthermore, Turkan et al. \cite{44-5} summarized the DL models for diagnosis of Alzheimer's disease based on the OCT and OCTA data. Study \cite{44-6} reviewed traditional statistical methods and deep learning models for detection, segmentation and quantification of neovasculariztion based on the OCTA data.

Compared to the aforementioned works, this survey focuses on the state-of-art DL models for OCTA data enhancement, image segmentation and volumetric data rendering, with the emphasis on the design of algorithms and model architectures. A systematic literature searching of peer-reviewed articles is performed. The main keywords used for literature searching are DL models, OCTA data, denoising, artifacts removal, segmentation and volumetric reconstruction. Articles published between January 1, 2020 and December 31, 2024 on PubMed, Scopus and Web of Science are considered. We found that the basic DL architectures including UNet, Transformer and Generative adversarial network (GAN) are most widely- adopted both in pre-2022 researches and in studies with a clinical application focus. Advanced DL models not only consist of these basic DL components, but also utilize the specific characteristics of the OCTA data. For example, the intensity distribution among consecutive B-scans, the signal attenuation physics, morphological similarities in the multi-scale vascular networks, angiogenesis dynamics are considered in the advanced DL models, so the performance of these models are greatly improved. Furthermore, to compensate for the limited amount of available OCTA data and to reduce the reliance on the label-intensive data annotations, the weakly-supervised and self-supervised training strategies of DL models are applied. In this survey, both the basic DL models and the novel and advanced model designs are introduced in details. The performance, strength and disadvantages of the models corresponding to various types of applications are compared. In the last part of this survey, the publicly available OCTA dataset, including the \textit{en face} images, B-scans and data volumes, are summarized. Overall, this review can help readers better understand various kinds of OCTA-based DL models, and assist potential users to select or design an appropriate model for data analysis in both fundamental research and clinical applications.

The rest of this paper is organized as follows. In Section II, three basic DL models, UNet, Transformer and GAN are introduced. These basic DL models with different loss function design are applied to segment large vessels in the OCTA \textit{en face} images, and their performances are compared. In Section III, the state-of-the-art (SOTA) DL models for denoising and enhancement of OCTA data, segmentation of the \textit{en face} OCTA images, and volumetric reconstruction of the fundus structures are discussed. Besides introducing the different design of model architectures, the supervised, weakly-supervised and self-supervised training strategies are discussed. The novel applications in the ophthalmology are summarized as well. In Section IV, eight publicly-available OCTA datasets are presented. Section V provides the conclusion and discussion.

\section{Basic DL Models and Training Strategies}
\label{Basic DL models}
This section introduces three basic types of DL models, UNet, attention-based models and GAN, for analysing OCTA data. These models are widely-used in practical applications due to their simple architectures, stable and good performance, less computational cost and open-source availability. In addition, the supervised, weakly-supervised and self-supervised training strategies of DL models are presented. The weakly-supervised and self-supervised training strategies are particularly useful for the small dataset in ophthalmology.\\

\subsection{UNet}

UNet is one of the most widely-used DL models for segmentation of biomedical images \cite{a1}. It contains a down-sampling path (the encoder) to extract latent states in relatively lower-dimension feature space, and an up-sampling path (the decoder) to reconstruct the segmented images using the latent features and input images. The basic building block in the encoder include the convolutional layer, normalization layer and nonlinear activation. In the decoder, each block contains a deconvolution operator, batch normalization and activation function. An example four-layer UNet structure is shown in Fig. 2a. 

Different types of loss functions, such as L1, mean square error (MSE), cross entropy (CE), DICE and Jaccard coefficient based losses, can be used for DL model training. For model evaluation, confusion matrix, structural similarity index (SSIM) and peak signal-to-noise ratio (PSNR) are widely used to evaluate the quality of prediction results, such as the removal rate of artifacts and segmentation accuracy, Generally, for an input image $i$, let the corresponding ground truth (target value) be $\mathbf{x}_i$ and the model prediction be $\hat{\mathbf{x}_i}$, where $i \in [1, N]$, $N$ is the batch size. Then, the most commonly adopted loss functions are summarized as follows:
 
 \begin{figure}[!t]
\centerline{\includegraphics[width=\columnwidth]{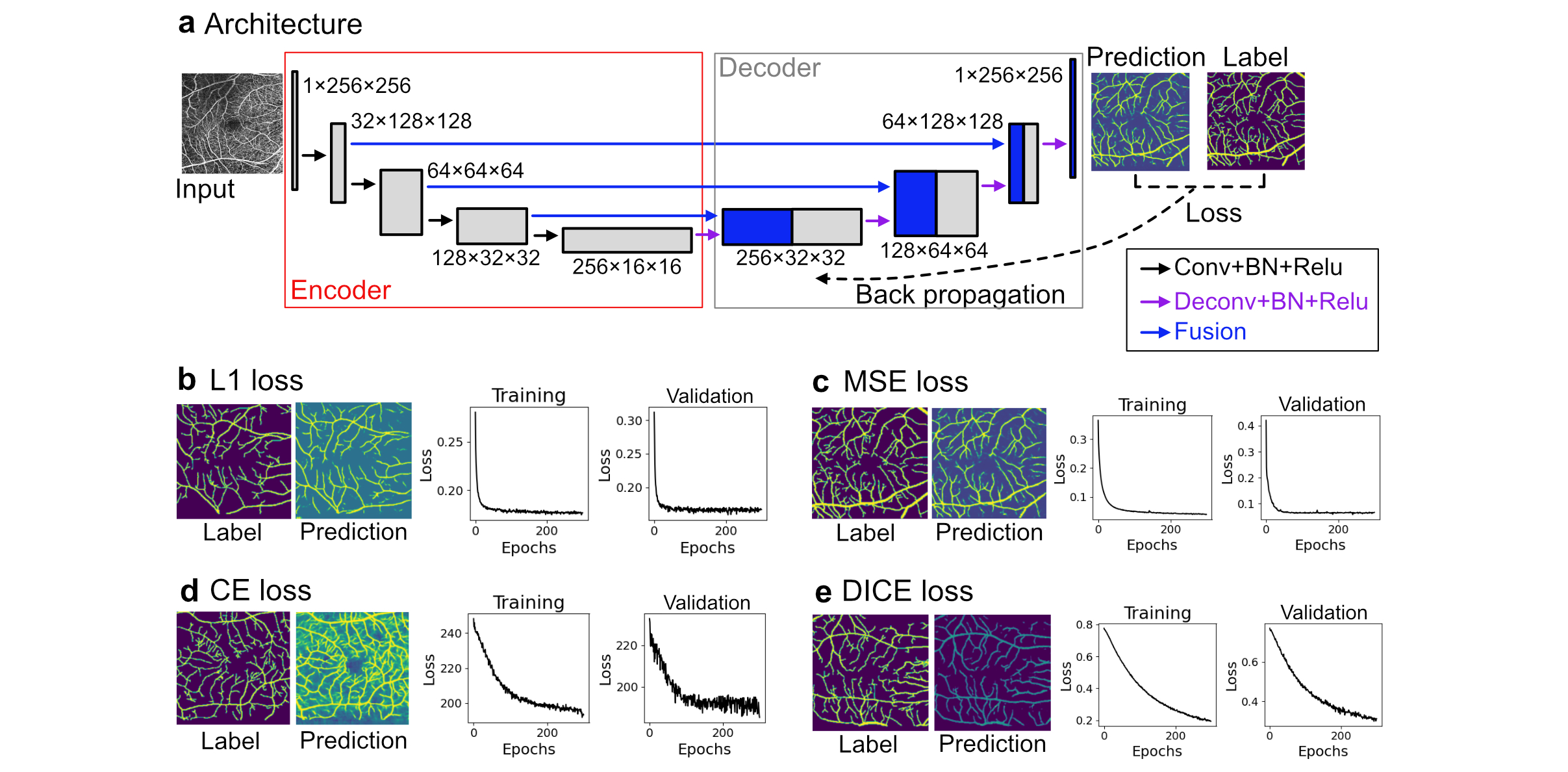}}
\caption{Architecture of UNets and vasculature segmentation results in OCTA \textit{en face} images. \textbf{(a)} Schematics of an example implementation of a four-layer UNet model. \textbf{(b-e)} Comparison of the segmentation performance and algorithm convergence of the UNet model when different loss functions are used, including \textbf{(b)} L1 loss, \textbf{(c)} MSE loss, \textbf{(d)} CE loss and \textbf{(e)} DICE loss.}
\label{fig2}
\end{figure}

\begin{itemize}
\item L1 loss \\
$L(\mathbf{x},\hat{\mathbf{x}}) = \frac{1}{N} \sum_{i=1}^N |\mathbf{x}_i - \hat{\mathbf{x}}_i|$. \\
\item MSE loss \\
$L(\mathbf{x},\hat{\mathbf{x}}) = \frac{1}{N} \sum_{i=1}^N (\mathbf{x}_i - \hat{\mathbf{x}}_i)^2$.\\
\item CE loss \\
$L(\mathbf{x}, \hat{\mathbf{x}}) = - \dfrac{1}{N} \sum_{N} \log \dfrac{\exp(\hat{\mathbf{x}}_t)}{\sum_j^K \exp(\hat{\mathbf{x}}_j)}$, where the label image $\mathbf{x}$ is used as the class indices, $t$ is the true class index, $j \in [1,K]$ is the indices of $K$ classes. \\
\item Jaccard coefficient-based loss \\
$L(\mathbf{x}, \hat{\mathbf{x}}) = 1-\frac{1}{N} \sum_{i=1}^N \dfrac{\mathbf{x}_i \hat{\mathbf{x}}_i}{\mathbf{x}_i+\hat{\mathbf{x}}_i - \mathbf{x}_i \hat{\mathbf{x}}_i}$, where $s$ is a smooth factor \\
\item DICE-based loss\\
$L(\mathbf{x}, \hat{\mathbf{x}}) = 1-DICE = 1 - \dfrac{1}{N} \sum_{i=1}^N \dfrac{2 \mathbf{x}_i \hat{\mathbf{x}}_i}{\mathbf{x}_i+\hat{\mathbf{x}}_i}$. 
\end{itemize}

The commonly used model performance metrics include:
\begin{itemize}
\item Confusion matrix\\
Most previous works cited in this survey used the confusion matrix to evaluate the performance of segmentation algorithms. Based on the confusion matrix, the following indexes can be calculated as\\
$Accuracy=\dfrac{TP+TN}{TP+TN+FP+FN}$, \\
$Sensitivity = \dfrac{TP}{TP+FN}$, \\
$Specificity = \dfrac{TN}{TN+FP}$, \\
$DICE = \dfrac{2TP}{FP+FN+2TP}$, \\
$False\,Discovery\,Rate = \dfrac{FP}{FP+TP}$,\\
$G-mean\,score = \sqrt{Sensitivity*Specificity}$,\\
where TP is the true positive, FP is the false positive, TN is the true negative, and FN is false negative. \\
\item SSIM 
\begin{align*}
SSIM = \dfrac{1}{N} \sum_{N} \dfrac{(2\mu_{x}\mu_{\hat{x}}+c_1)(2\sigma_{x\hat{x}}+c_2)}{(\mu_x^2+\mu_{\hat{x}}^2+c_1)(\sigma_x^2+\sigma_{\hat{x}}^2+c_2)}.
\end{align*}
where $\mu_x$ is the average intensity of $\mathbf{x}$, $\sigma_x^2$ is the intensity variance of $\mathbf{x}$, and $\sigma_{x\hat{x}}$ is the covariance of $\mathbf{x}$ and $\hat{\mathbf{x}}$. $c_1$ and $c_2$ are constants influenced by the size of $\mathbf{x}$. SSIM index is usually used to evaluate the algorithm performance for OCTA data denoising and enhancement.\\

\item PSNR \\
$PSNR = \dfrac{1}{N} \sum_{i=1}^N \log_{10} \big(\dfrac{MAX^2}{MSE_i} \big)$, where $MAX$ is the maximum value of the image intensity, and $MSE_i$ is the mean square error between the $i^{th}$ label and prediction.\\
\end{itemize}

We use a toy example to show the training procedure using the UNet structure given in Fig.~\ref{fig2}a for segmentation of vessels. Consider that 106 OCTA \textit{en face} images of DR disease and their corresponding annotations of vessels are rotated and flipped to generate 212 image data. Out of them, 200 images are used for model training and 12 images are for validation. The stochastic gradient descent solver ADAM (learning rate=0.001, $\beta_1=0.9$, $\beta2=0.999$) is used for model training. The small data size and the limited labels influence the model performance significantly, but this is a common problem in many ophthalmic applications.

After 300 epochs, the four-layer UNet model is able to capture large vascular features, while some small vessels cannot be recognized and segmented correctly. There are also some fragmented vascular pieces in the predicted images. Increasing the model complexity, such as the depth of the neural network, can potentially improve the capability of extracting small-scale features as well as the connectivity of segmented vessels. Results also show that models with L1 and MSE losses can converge very fast in less than 50 epochs (Fig. 2b and c). The training of CE loss-based model is no as stable as using other loss functions. The DICE losses keep decreasing in the 300 epochs, but the model suffers from an over-fitting problem after approximately 200 epochs. Thus, the predicted images do not have distinguishable improvements during the 200-300 epochs. Furthermore, models with MSE and CE losses are likely to capture more small-scale features based on the same model architecture.

In addition, if the upsampling blocks in the decoder only uses the features from the previous block, it is a basic CNN-based autoencoder. For example, in Fig.2, an autoencoder only has information propagating through the black and purple arrows. But in many situations, fusion of the low-level features extracted by the encoder blocks and the upsampled features, represented by the blue arrows, greatly improves the inference power. Thus, the UNet architecture is more frequently used than a basic autoencoder, especially for image segmentation task.

\subsection{Attention-based Models}
Attention-based models are typically categorized into two architectural types: self-attention mechanisms (e.g., Transformers)~\cite{a2,a3,a4} and spatial attention mechanisms~\cite{a5,a5-2,a6}. In UNet-based models, attention mechanisms are commonly integrated at the bottleneck between the encoder and decoder to weight the importance of individual elements or local patches within feature maps~\cite{a2,a3,a4,a5,a6}.

\begin{figure}[!t]
\centerline{\includegraphics[width=\columnwidth]{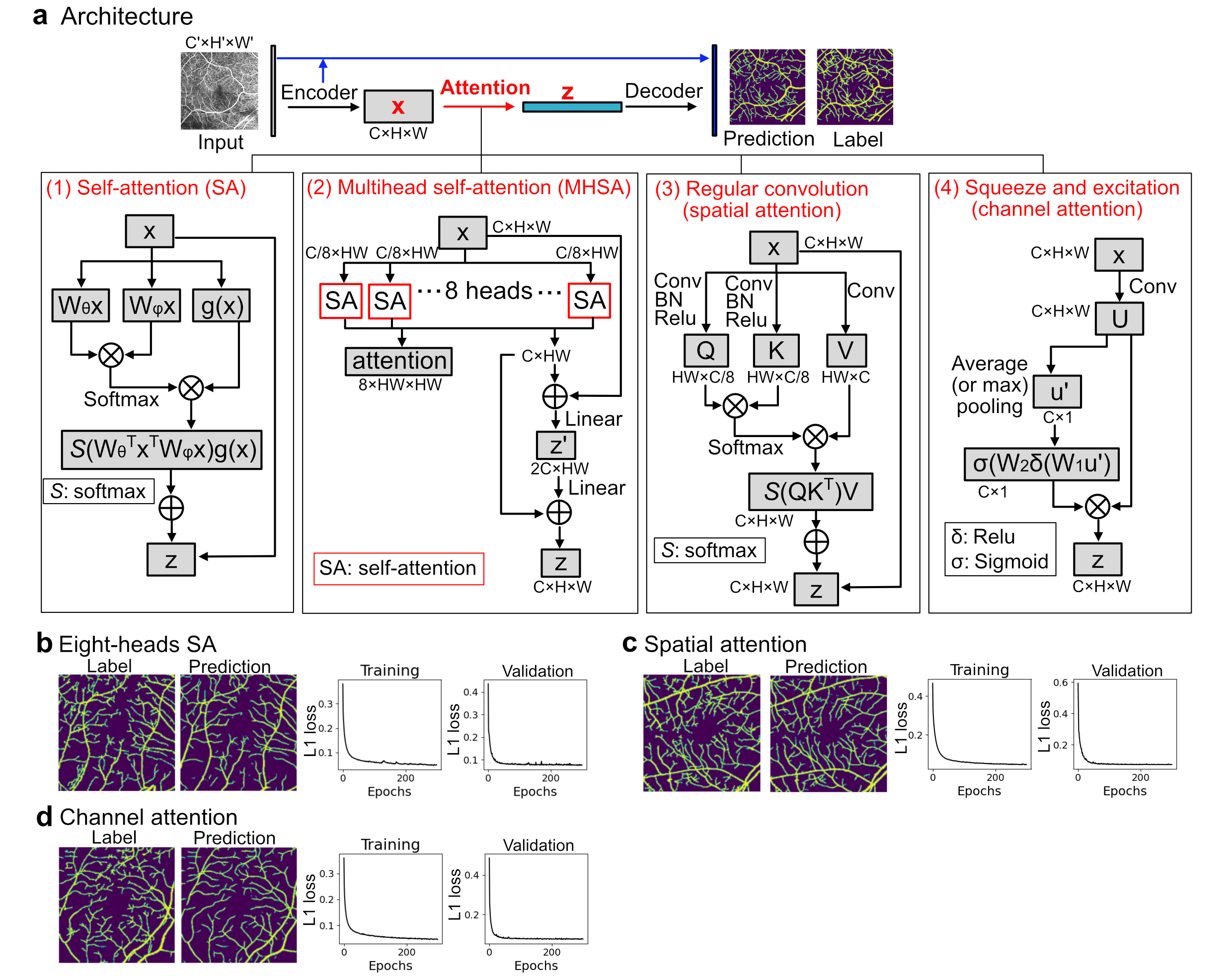}}
\caption{Architecture of typical attention mechanisms and vasculature segmentation results in OCTA \textit{en face} images. \textbf{(a)} Schematics of an example implementation of different attention-based models, which have a UNet structure as the backbone model. The attention blocks are placed at the bottle neck of the encoder and decoder. The structures of the attention block, which has four variants as self-attention, multihead self-attention (MHSA), spatial attention and channel attention. \textbf{(b)} Comparison of the segmentation performance and algorithm convergence of the Transformer models with different attention blocks, including \textbf{(b)} MHSA, \textbf{(c)} spatial attention and \textbf{(d)} channel attention. L1 losses were used.}
\label{fig3}
\end{figure}

Among various attention mechanisms, the major breakthrough architecture is the self-attention (SA), where the elements or patches in the input tensor attend to all others in the same tensor \cite{a3,a4}. As such, the global dependencies are captured. The output of a SA block is calculated as the dot product of the attention weights $A = Softmax(\mathbf{W}_{\theta}^T \mathbf{x}^T \mathbf{W}_{\phi} \mathbf{x})$ and the linear embeddings of $\mathbf{x}$, $g(\mathbf{x})=\mathbf{W}_g \mathbf{x}$. $\mathbf{W}_{\theta}$, $\mathbf{W}_{\phi}$ and $\mathbf{W}_g$ are learnable weigh matrices updated during model training. Fig. 3a shows one possible implementation of a SA block. 

Multihead self-attention (MHSA) block utilizes different features, such as features at multiple spatial ranges extracted by multiple self-attention heads \cite{a3}. Thus, the model can attend different representation subspaces. The MHSA block is the foundation of the Large Language Models (LLMs) such as BERT \cite{a7}, GPT \cite{a8}, etc. Fig. 3a shows the schematics of one possible implementation of MHSA.

Spatial attention \cite{a5} and its variant, channel attention \cite{a6}, encode spatial positions and activations/deactivations of channels in the attention contents. For example, the convolution-based spatial attention (Fig. 3a) has a similar structure as the SA block. But in a spatial attention block, the convolution operators are applied to the input tensor $\mathbf{x}$, such that the positional information is included in the attention factors Q, K and V \cite{a5}. The channel-wise attention block uses global average or max pooling to generate channel-wise statistics in the \textit{squeeze} operation. Then, based on these statistics, channels are selectively activated or deactivated using a gating mechanism with a sigmoid activation function, named as the \textit{excitation} operation. Thus, the output of the block re-allocates the importance to the most informative channels of an input image or tensor \cite{a6}.

The performance of the UNet-based DL models with different attention blocks was quantified. A UNet which has five convolution blocks as the encoder, four convolution blocks and four deconvolution operators as the decoder was used as the backbone model. The backbone model is first proposed in CSNet model \cite{a9}. The attention blocks between the encoder and decoder were eight-head SA, convolution-based spatial attention and channel attention, respectively. Results in Fig. 3(b)-(d) show that all the three Transformer models can converge with 100 epochs, where the L1 losses are smaller than the basic UNet model in Fig. 2b. For the three Transformer models in Fig. 3(b)-(d), the L1 losses of the validation data at the $300^{th}$ epoch were 0.076, 0.074, 0.076, respectively. The L1 loss of the basic UNet model was 0.17. Though some small-scale structures still cannot be captured correctly by the Transformer models, the overall shape of the vasculature and the connectivity of vessels are better than the UNet. Besides, the performance of the spatial attention-based Transformer is slightly better than the other two models, where it recognizes more micro-vessels at the bifurcating ends of the vasculature networks. 

\subsection{GAN}
The core component of a GAN is a generator (typically a CNN-based autoencoder~\cite{a10} or a UNet~\cite{a11}) that maps latent variables to synthetic samples with a learned probabilistic distribution. Adversarially, a discriminator network (often another CNN-based model) is trained to distinguish between generated samples and real data. The discriminator minimizes its classification error via binary cross-entropy loss, while the generator aims to deceive the discriminator by producing samples indistinguishable from real data, thereby maximizing the discriminator’s error rate. 

The generator and discriminator networks, compete with each other during the dynamic model training and thus form a zero-sum game, which enables the generator to produce more accurate and realistic predictions compared to the target data. GAN is one of the most widely-used generative models for image-to-image translation \cite{a9,a10,a11}. It has many variants such as deep convolutional GAN (DCGAN) \cite{a12}, perceptual GAN \cite{a13} and cycle-consistent GAN (CycleGAN) \cite{a14}. 

\begin{figure}[!t]
\centerline{\includegraphics[width=0.9\columnwidth]{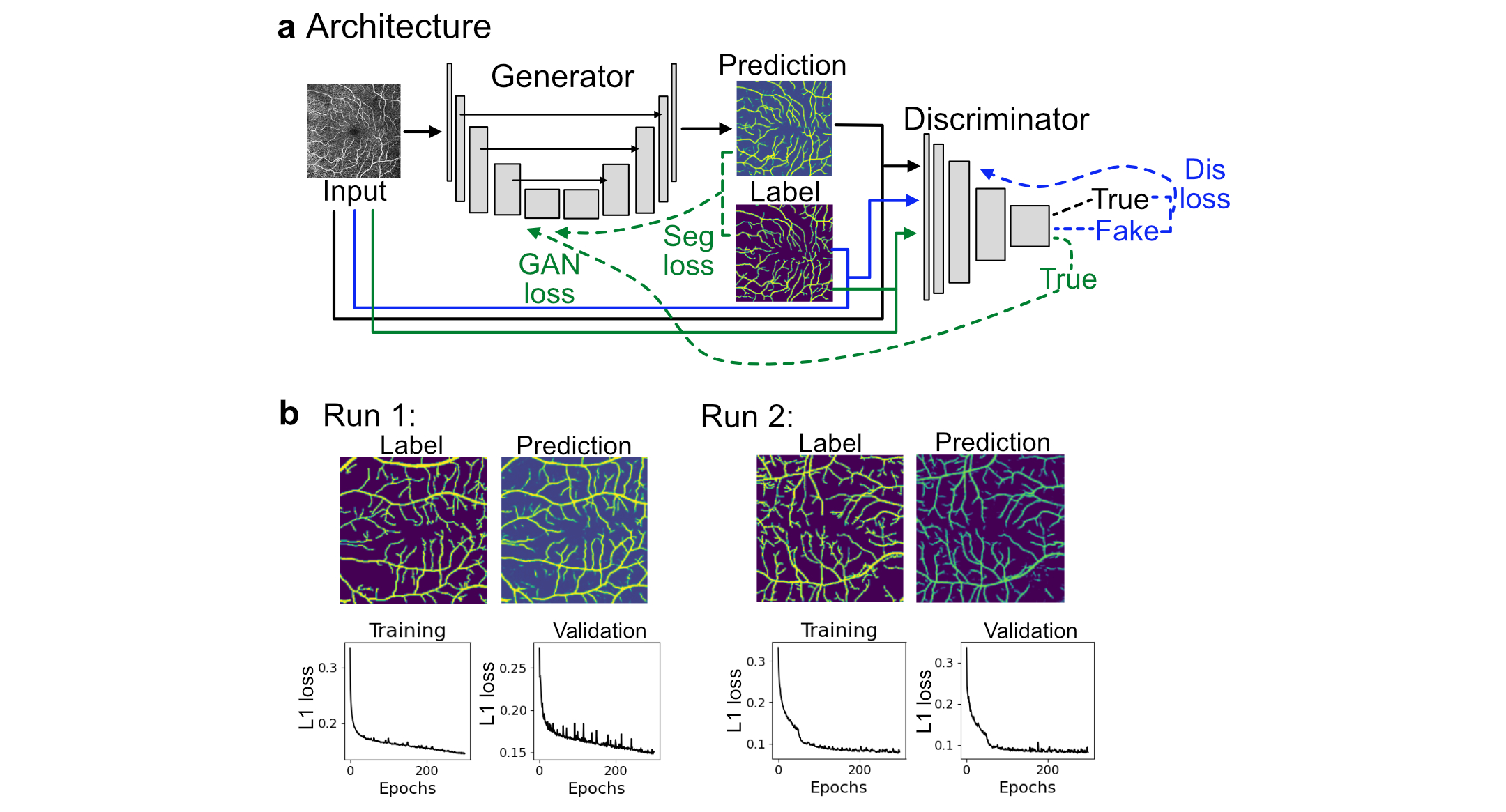}}
\caption{Architecture of GANs and vasculature segmentation results in OCTA \textit{en face} images. \textbf{(a)} Schematics of an example implementation of a GAN model, which has a UNet as the generator and a CNN as the discriminator. \textbf{(b)} Comparison of the segmentation performance and algorithm convergence of the GAN model at two training runs.}
\label{fig4}
\end{figure}

A basic GAN model for vasculature segmentation is shown in Fig.~\ref{fig4}a. This GAN model adopts a four-layer UNet model as the generator and a five-layer CNN as the discriminator.The generator and discriminator networks are trained in tandem in each epoch. To train the generator, the segmentation loss (Seg loss) between the predictions and true labels is calculated and propagated back-through the generator. At the same time, the discriminator loss (Dis loss), indicating whether the discriminator is fooled by the generated predictions, is also propagated back-through the generator with a loss weight 0.03. Meanwhile, training of the discriminator network relies on propagating the Dis loss back-through the discriminator. The Dis loss quantifies the ability of the discriminator to distinguish the differences between the predictions and true labels. Results in Fig.~\ref{fig4}b show that the segmentation accuracy of the generator implemented in a UNet structure, is significantly improved in the GAN model compared with the basic UNet model in part (1). The L1 losses of the GAN model are 0.151 and 0.082 in two training rounds where the model parameters are initialized with different random numbers, respectively.

It should be noted that though the adversarial mechanisms between the generator and the discriminator of a GAN model can greatly improve the prediction power of the generator, the convergence of the training is not guaranteed and often collapses into failure modes. For example, the imbalance between the generator and discriminator due to different learning rates of the two networks is a common cause of the mode failure. As one possible outcome of the training, a strong discriminator and a weak generator will not be able to make correct predictions. 

\section{DL Models for OCTA Data Enhancement}
The quality of OCTA data is largely influenced by the acquisition artifacts, inherent noises, and inter-device variances. Any spurious alterations in the OCTA data can lead to misinterpretations of the anatomical and pathological structures, and influences clinical diagnosis accuracy \cite{17,18,19,20,21,22,23,24}. In this section, the state-of-the-art DL models for (a) removal of artifacts, (b) denoising and enhancement and (c) removal of inter-device variations are introduced.

\subsection{Removal of Artifacts}

\begin{figure}[!t]
\centerline{\includegraphics[width=0.7\columnwidth]{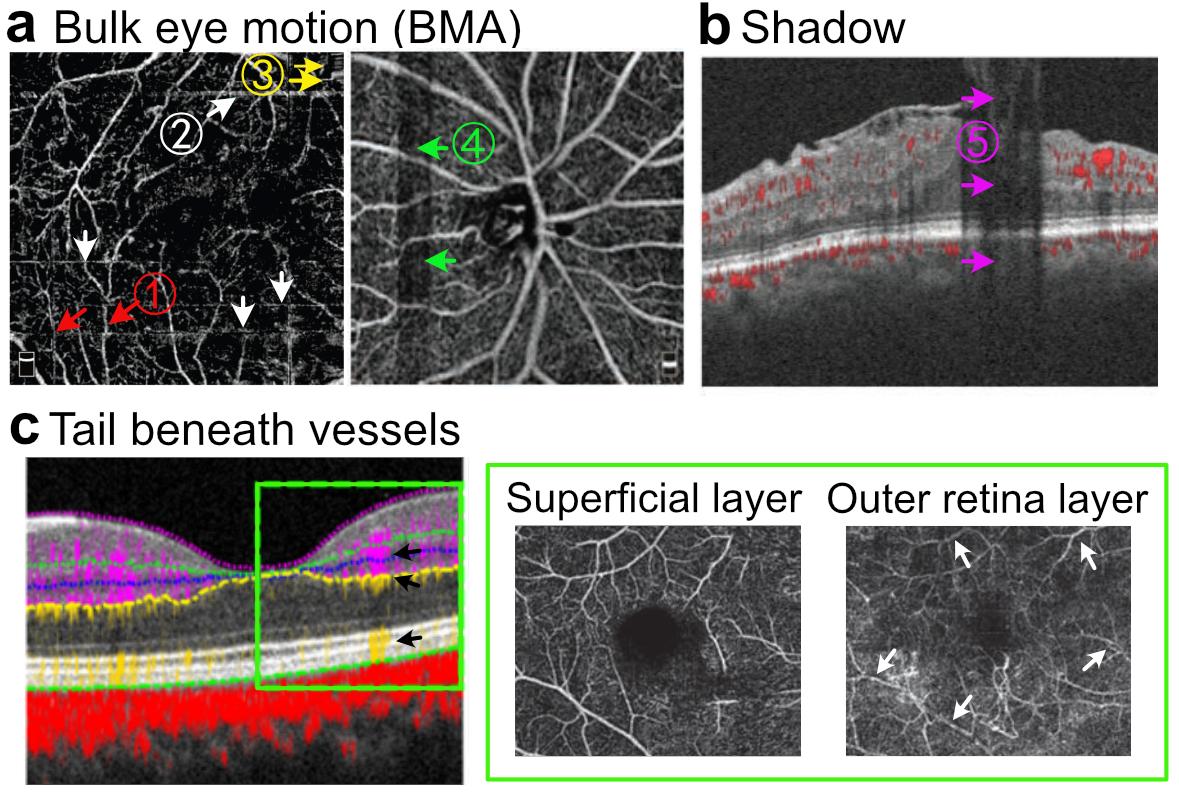}}
\caption{Examples of common artifacts in the OCTA data. \textbf{(a)} Bulk eye motion artifacts (BMA). In the \textit{en face} OCTA images, there are \textcircled{1} discontinued vessels, \textcircled{2} stripes, \textcircled{3} stretching and \textcircled{4} low-signal bands \cite{18}. \textbf{(b)} Shadow artifacts due to the light blockage. which creates a low-signal band \textcircled{5} under the local blockage in a B-scan image \cite{18}. \textbf{(c)} Artificial tail signals beneath the large vessels lead to the ghost vessel artifacts. Example tail signals are marked by the black arrows in the B-scan image. Ghost vessels in the \textit{en face} image of the outer retina layer are generated, marked by the white arrows \cite{f1,18}.}
\label{fig5}
\end{figure}

Large eye motion due to the movements of patient's eye, head and body leads to distinguishable and widespread deccorelation over the B-scan images, which can be automatically removed by most commercial OCTA equipment \cite{1,18,20}. However, high frequency eye movements and pulsatile fluctuations of retina and choroid are hard to be detected and corrected by these commercial software. Thus, stripes, bands with various brightness, discontinuity or duplication of vessels are generated (Fig. 5a) \cite{18}. Shadow artifacts are caused by the defocus of scan beam, projection attenuation and local blockage \cite{1,18,23}. For example, Fig. 5b shows a vertical low-signal band generated due to the local light blockage by superficial edema or vitreous floaters \cite{18}. Artificial signals beneath large retinal vessels are another common source that lead to the ghost vessel artifacts (Fig. 5c) \cite{18}. This projection artifacts of overlying retina features in the deep image layers is one of the most important artifacts that cause ambiguity and misinterpretation of retinal structures and pathologies \cite{24}.

Automatic removal of artifacts can be challenging, because of their irregular shapes, large variations of the intensity and random distributions in the data. Advanced DL models were developed to improve the capability and efficiency of detecting and removing artifacts. The advanced DL models were constructed by stacking multiple basic DL models in a series or parallel manner. The composition blocks of the backbone models, such as the convolution blocks, were replaced by advanced designs. Novel loss functions and training strategies were applied to improve the performance \cite{45,46,47,48,49,50,51,52,53,54,55}. 

\begin{figure}[!t]
\centerline{\includegraphics[width=\columnwidth]{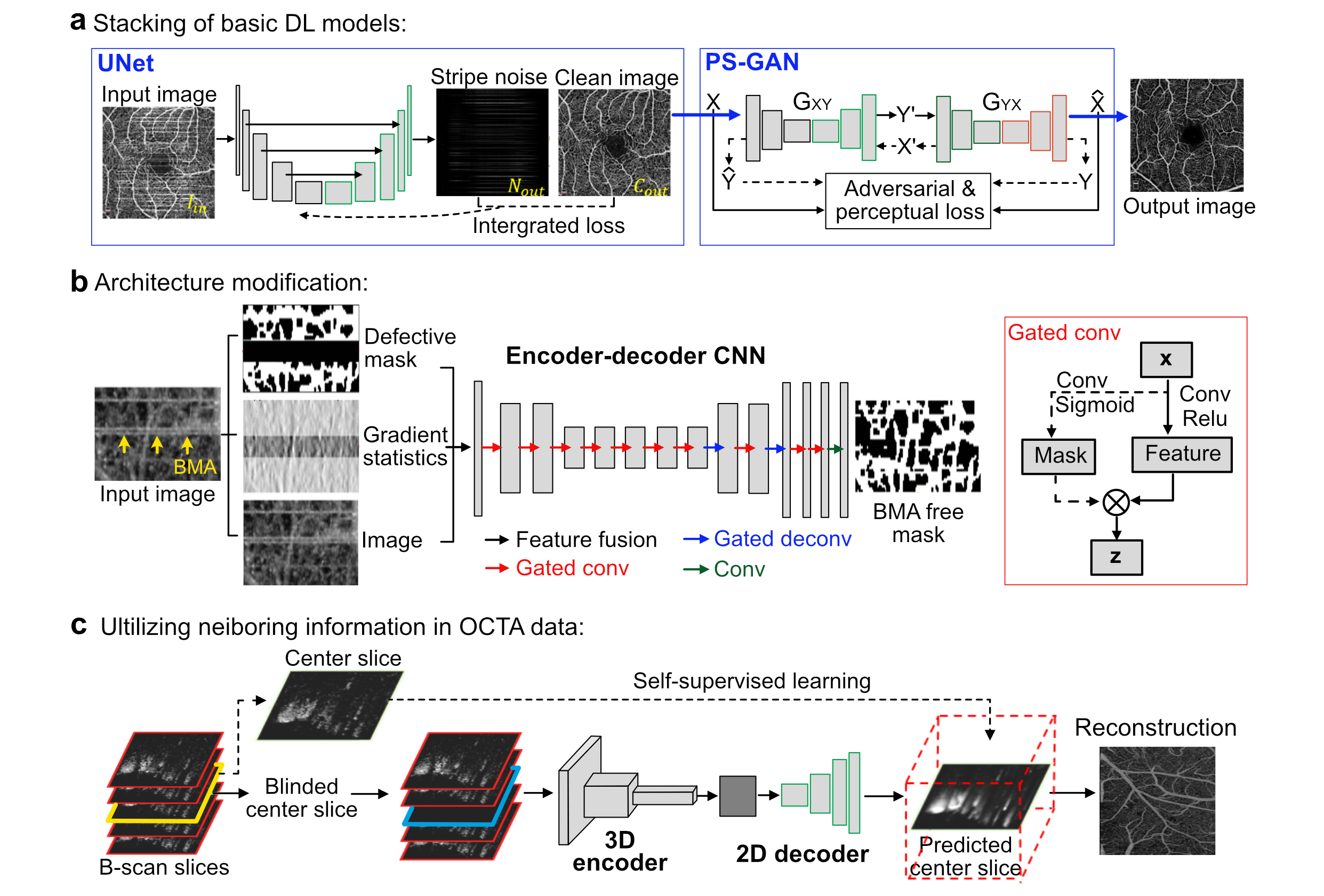}}
\caption{The architectures of three representative DL models for removal of artifacts. \textbf{(a)} An integrated DL model for removing stripe noises in the \textit{en face} images by stacking the UNet and GAN models in a series manner \cite{48}. \textbf{(b)} An CNN-based autoencoder model for removal of bulk motion artifacts (BMA) by using gated convolution and deconvolution as base blocks and fusion of multiple appearance features as model input \cite{47}. \textbf{(c)} An CNN-based autoencoder model for desnoising, which is trained based on a self-supervised learning strategy using the neighboring information in the data volume \cite{51}.}
\label{fig6}
\end{figure}

For example, in~\cite{48}, a five-layer UNet model was used to decompose the raw OCTA images to the clean images and stripe noises. An integrated loss function was designed, which contained the reconstruction loss, stripe loss and ATV loss. The construction loss measured the difference between the input images and the output images and noises. The stripe compared the output noises and the reference stripe noises. ATV loss was for evaluating the completeness of vessel structures. Then, the perceptual structural GAN (PS-GAN) was developed for enhancement of small-scale vessel structures. The backbone architecture of the PS-GAN was a CycleGAN, which consisted of two symmetric encoder-decoder networks to translate an input image from source domain $X$ to a conditional domain $Y$, and finally to the target domain $\hat{X}$. Meanwhile, latent data in the conditional domain $Y$ was back-propagated through the two networks. The adversarial and perceptual losses of $X$ and $\hat{X}$, $Y$ and $\hat{Y}$ were calculated for the training process (Fig. 6a). Thus, by stacking the two basic models in a series manner, the removal of stripe noises and image enhancement were achieved. Similar models constructed by integrating a few basic models can be found in \cite{46,50,52,53,54,55}.

In~\cite{47}, the CNN-based autoencoder was used as the backbone model, where its base blocks was replaced by the gated convolution and deconvolution blocks. A gated block \cite{57} has two information propagation paths. As shown in the red box in Fig. 6b, the solid lines indicate a basic convolution path for extracting spatial features, and the dashed lines is the branch that identifies the stripe areas using a kind of attention convolution mechanism. Moreover, information fusion and injection was applied to better leverage the structural and texture information in the BMA-affected areas. The defective mask, in which the BMA-affected areas were removed, was generated using the optimally oriented flux method \cite{56}. The horizontal gradients were extracted by applying the vertical Sobel operator to the raw image \cite{47}. The mask, gradient statistics and images were combined and inputted into the DL model, such that features in the BMA-affected areas were inpainted. Thus, this type of DL models uses a regular encoder-decoder architecture but modifies the information propagation paths and mathematical operators in the network. Another work \cite{45} modified the information propagation paths in a basic UNet model and inputted the volume of OCT time-series to construct the clean OCTA images.

The third representative model can be found in~\cite{51}. Based on the assumptions that vascular structures are spatially continuous, noises are uncorrelated and stripe areas have different brightness, a DL model, SOAD, was developed for denoising the OCTA B-scan data by masking partial information. As shown in Fig.~6c, the backbone of SOAD is an autoencoder which takes the volume of five spatially consecutive B-scan slices as the input. The center slice in the input data volume is masked as black. The encoder-decoder network predicts information in the center slice, and then prediction results are compared with the original raw images in a self-supervised training process. This method is applied to remove the shadow artifacts beneath the large vessels in the B-scan slices. Afterwards, the vascular structures in the reconstructed OCTA \textit{en face} images are enhanced and many ghost vessels and noises are removed. Other self-supervised training mechanisms can be found in the work using artifact-free areas in the same images \cite{47} and using noisy OCTA images generated based on the correlation analysis of B-scans as the ground truths \cite{45,49,50} for the training process.

\subsection{Denoising and Data Enhancement}

The inherent noise in the OCT B-scans comes from the equipment and measurement protocols, which is hard to be separated from the true signals. Acquiring a large number of consecutive B-scans at the same location for the correlation analysis could help reduce the noise and blurry in the generated \textit{en face} OCTA images. However, it would require greatly increased scanning time and higher scanning speed \cite{1}. When the vasculature of animal brains was studied in the laboratories, more than 48 consecutive B-scans can be acquired to improve the data quality. However, in clinics, the commercialized OCTA equipment usually allows no more than 8 consecutive B-scans of human eyes. Thus, it is important to develop novel models for removing noises and enhancing the quality of OCTA data.

\begin{figure}[!t]
\centerline{\includegraphics[width=\columnwidth]{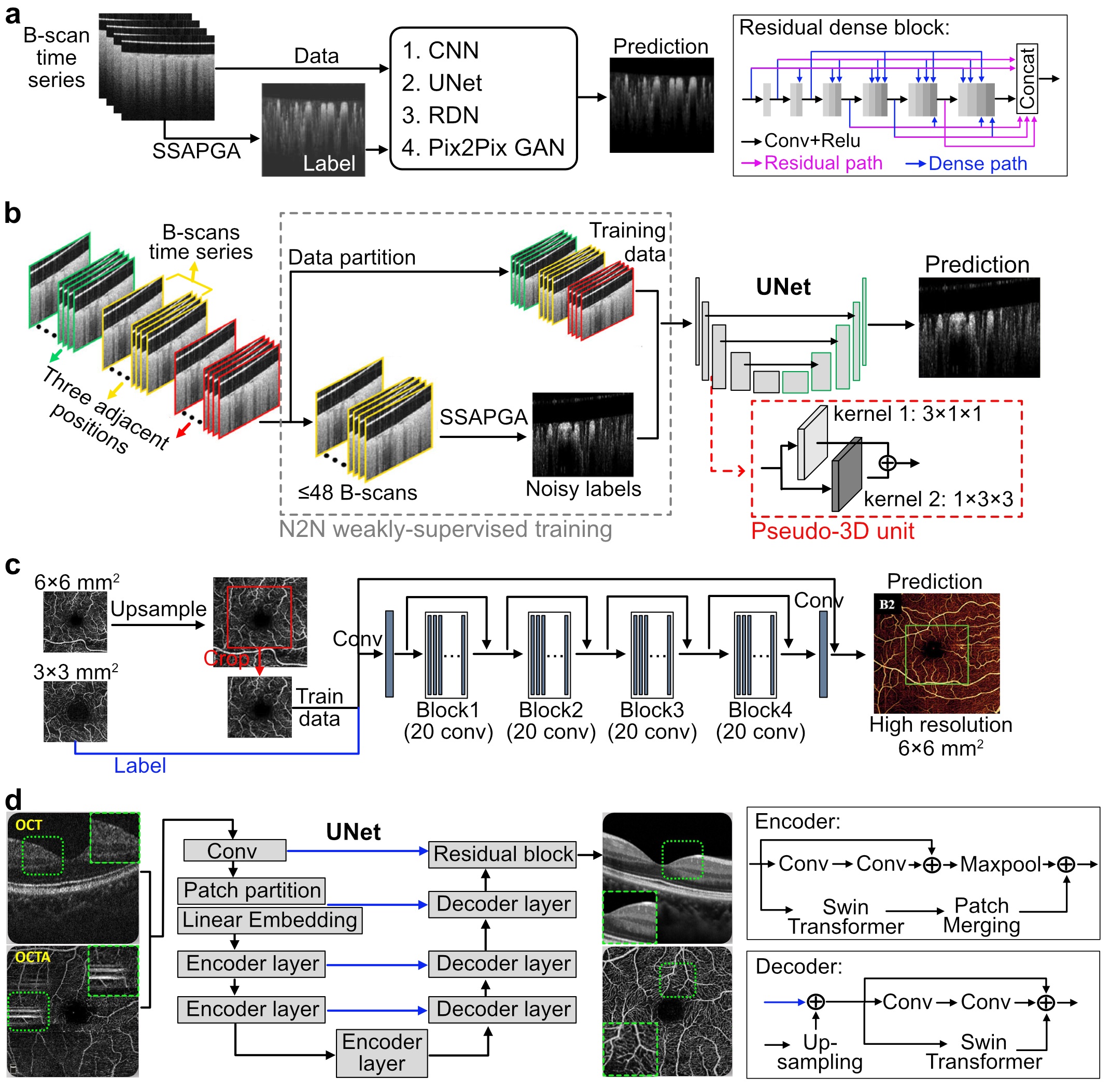}}
\caption{The architectures of four representative DL models for OCTA data denoising and enhancement. \textbf{(a)} Four basic models for removing the noises in the OCTA B-scan slices. The RDN has an encoder-decoder structure with residual dense blocks \cite{59}. \textbf{(b)} A UNet model which takes the spatial- and temporal-varying OCT B-scan slices and predicts the noise-free OCTA B-scans. N2N weakly-supervised training strategy and the pseudo-3D units for extracting the spatial and channel-wise features is applied \cite{66}. \textbf{(c)} A residual dense CNN for enhancement of OCTA image resolution. The model uses $3 \times 3 \, mm^2$ high-resolution \textit{en face} images as the training labels and reconstructed the $6 \times 6 \, mm^2$ images of a large field view as well as high resolution \cite{61}. \textbf{(d)} A UNet model integrated with swin transformer for denoising the OCT and OCTA images \cite{69}.}
\label{fig7}
\end{figure}

In the last five years, there are fourteen papers about the advanced DL models for denoising and data enhancement \cite{51,58,59,60,61,62,63,64,65,66,67,68,69}. Some of these works also achieved removing the artifacts \cite{51} or segmentation of anatomical structures \cite{58,69}. For example, in~\cite{59}, the authors compared the performance of four basic DL models on denoising the OCT B-scan time-series and reconstructing the OCTA data (Fig. 7a). The structures of CNN, UNet and GAN have been introduced in Section 2. The RDN model contains 20 residual dense blocks, where each block uses residual paths and dense paths to propagate the hierarchical features as shown in the black box in Fig. 7a. The RDN model achieves a better performance compared with other three models by utilizing the multi-scale features. Studies \cite{58,60,68} used basic UNet models for data denoising and down-stream clinical applications, such as the quantification of the disease-relevant features of the foveal avascular zone (FAZ) and vessels. Two other studies developed multiple feature extraction paths, which were aligned in a parallel \cite{62} or a series manner \cite{65}, in the CNN model. The study \cite{67} applied a basic GAN model to denoise the OCTA images of the anterior segment vessels. Then, the vascular characteristics such as the vessel density and vessel diameter index of patients' data were quantified, enabling the accurate differentiation of the glaucoma severity. 

Two studies \cite{63,66} applied the Noise2Noise (N2N) weakly-supervised training strategy for the UNet model (Fig.~\ref{fig7}b). It has been proved mathematically that a neural network can be trained to generate high-quality clean data, by using a large amount of noisy data without labels and a small amount of noisy data with noisy labels \cite{70}. Thus, in these studies, the data was partitioned into two groups, where most noisy OCT B-scans were directly inputted into the UNet model and a small number of noisy data were used to generate noisy OCTA images as labels based on the correlation analysis method (SSAPGA). In addition, in~\cite{66}, both spatial and temporal-varying OCT B-scans were used as the training data. The base convolution block of the UNet model was modified as a pseudo-3D unit which performed both spatial and channel-wise convolution operations to extract features.

In~\cite{61}, deep residual network was developed for reconstruction of high-resolution OCTA \textit{en face} images. To prepare the training data, both high-quality $3 \times 3 \, mm^2$ OCTA \textit{en face} images and low-resolution $6 \times 6 \, mm^2$ with a large field of view were acquired in the experiments. These two image types had the same sizes, but a larger area of retinal vasculature was captured by the $6 \times 6 \, mm^2$ images. The $6 \times 6 \, mm^2$ images were firstly upsampled, and their center areas which had the same vascular structures as the  $3 \times 3 \, mm^2$ images were cropped to generate the training data. The corresponding high-quality $3 \times 3 \, mm^2$ images were used as the paired labels (Fig. 7c).

In the study of \cite{69}, the base component of a UNet model was modified by using both convolution and transformer operators to extract informative features. As shown in Fig. 7d, the swin transformer was used as an additional path to extracted features in both encoder and decoder layers. The swin transformer was a modified MHSA (Fig. 3a (1)) which computed the self-attention in the small windows instead of performing the global computation. There were two kinds of swin transformer, the window-MHSA and shifted window-MHSA. With window-MHSA, the attention windows were designed to partition the images in a non-overlapping manner. In order to consider the information correlation between different windows, the positions of the attention windows were spatially shifted in the second type \cite{71}. Overall, the UNet model and many of its variants were the most widely-used in denoising and enhancement of OCT and OCTA data. 

\subsection{Removal of Inter-device Variations}

Due to the large variations of technical specifications of commercial OCTA devices, the inter-device differences in the OCTA data is influential to the diagnostic results. Several comparative studies have highlighted the inconsistencies in the information obtained from different OCTA devices \cite{72,73,74}. Specifically, a commentary published as~\cite{74} summarized the effects of inter-device variations in OCTA data and described how to apply generative models such as CycleGAN and diffusion models~\cite{yang2023diffusion} for addressing this issue.

Previous works based on CT, MRI and PET data have successfully demonstrated that generative models can be trained to efficiently translate the data from a source domain to a target domain \cite{75,76}. Several recent works have achieved image translation from OCT to OCTA \cite{77,78,79}. However, very few models have bee proposed to solve the inter-device variations of OCTA data, because of the practical challenges including the large inter-device differences, small dataset but a broad spectrum of anatomical structures of disease eyes, large computational cost and the intrinsic difficulty of training a generative model.  

Recently, the study in~\cite{80} discussed the possibility of reconstructing OCTA images using the volumetric OCT data and then transferring the OCTA generator trained on one device to another device. The proposed model used a basic GAN architecture as described in Section II as the backbone model. Two context-enhanced encoders, two domain-specific OCT decoders and one domain-invariant OCTA decoder were constructed. The paired OCT and OCTA data was used to train the source-domain OCT encoder and the OCTA decoder using the adversarial training mechanism. Then, the two domain-specific OCT encoder-decoders for translating OCT data from target domain to source domain were constructed and trained based on the unpaired OCT data acquired from two devices, using the adversarial loss, cycle-consistency loss and reconstruction loss. This work provided inspiring ideas for designing new models to address the inter-device variations and improve the homogeneity and compatibility among various OCTA dataset.\\

\section{DL Models for Segmentation}

There were many more works on designing segmentation models for the anatomical structures in the OCTA data, such as the vasculature \cite{81,84,87,96,111,112}, foveal avascular zone (FAZ) \cite{107}, non-perfusion areas (NPAs) \cite{85,55}, retinal fluid volumes \cite{89,115} and neovascularization areas (NV) \cite{82,82-1,97-1,88}. We found more than forty papers published in the last five years, developing novel DL models for segmentation of OCTA data. In this section, a comprehensive review about these models is provided from three perspectives: the representative model architectures, weakly-supervised and self-supervised training strategies for reducing tedious annotation works, and their practical applications.  

\subsection{Representative Model Architectures}
\label{sec-4-1}
The earlier works at around 2020 focused more on supervised methods using basic CNN and UNet models \cite{81,124,82,83,84,85}. These models took the \textit{en face} OCTA images and pixel-wise annotations of vessels from the experts as input, and predicted the binary segmentation results. In~\cite{81,124}, the performance of the basic UNet (Fig. 2a) and CSNet (Fig. 2a and c) models were compared with traditional filter-based methods such as Frangi filter, Gabor filter and optimally oriented flux (OOF). The neural network models were shown to achieve the best segmentation results. 

\begin{figure}[!t]
\centerline{\includegraphics[width=\columnwidth]{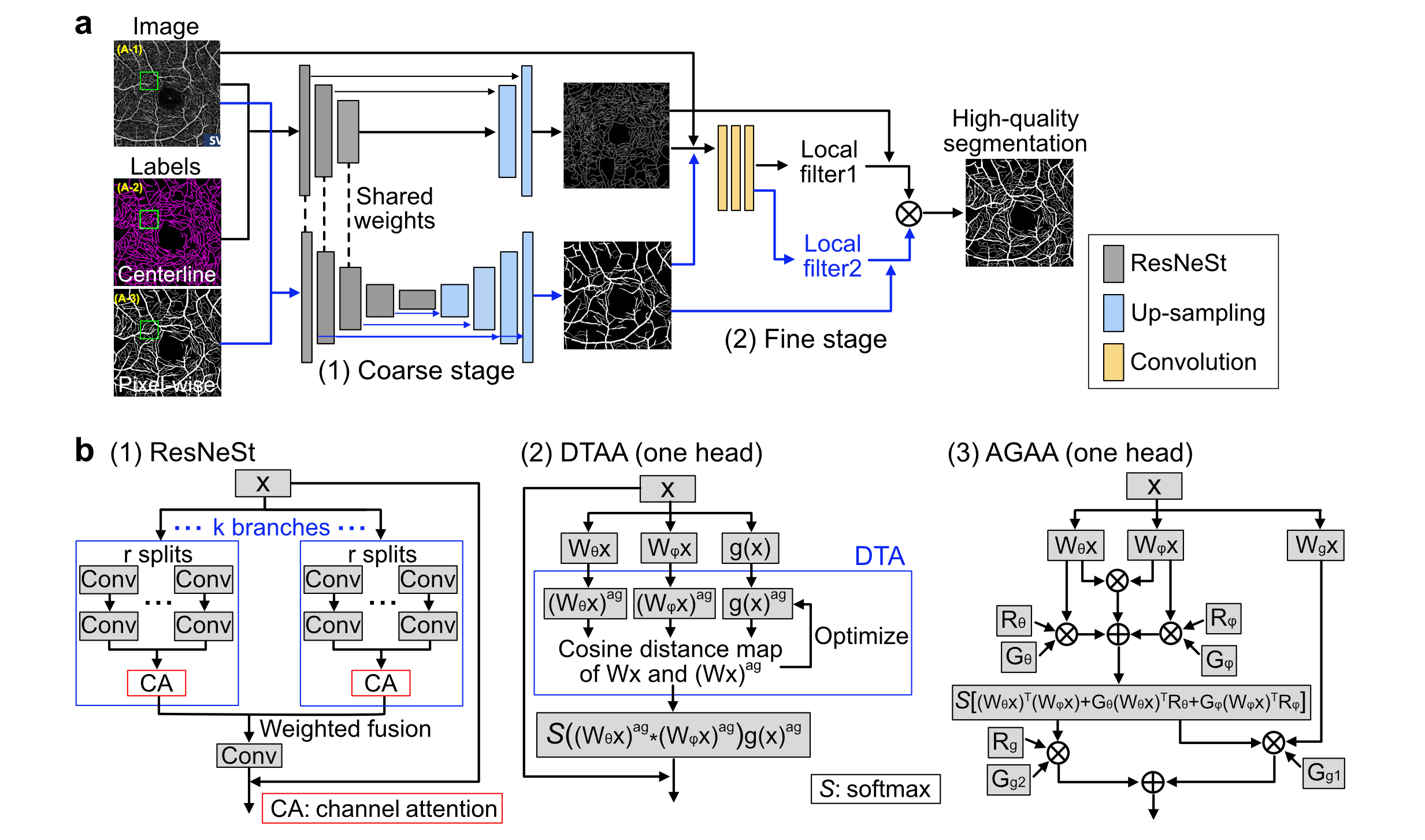}}
\caption{The representative architectures of segmentation models. \textbf{(a)} The dual-branch model, OCTA-Net, for segmentation of both large-scale arteries and veins as well as microvasculature \cite{90}. \textbf{(b)} Three novel attention mechanisms used in the transformer blocks. (1) ResNeSt used in \cite{90}. (2) Dynamic token aggregation attention (DTAA) used in \cite{100}. (3) Adaptive gated axial attention (AGAA) used in \cite{102}. Note that, (2) and (3) only show the structure of one head, but multiple heads were used in the transformer blocks in these two studies \cite{100,102}.}
\label{fig8}
\end{figure}

Many novel and clinical-relevant applications can be investigated using the basic UNet structure \cite{124,82,82-1,83,83-2,83-3,86,87,55,88,115}. For example, the study of \cite{82} used two UNet models, one for identification and segmentation of the choroidal neovascularization (CNV) areas and another for segmentation of micro-vasculature in the CNV. The study \cite{83} inputted both OCT and OCTA \textit{en face} images into a UNet model to segment and differentiate arteries and veins. Beside identifying the large vessels, the UNet model can also be applied to separate areas filled with arterial or venous micro-vasculature in the OCTA \textit{en face} images \cite{83-2,83-3}. Moreover, the study in \cite{55} designed a DL model by integrating four UNet models in a parallel way. The first two parallel UNets extracted features from the thickness map of the inner retina, OCT reflectance image and OCTA image. Then, these features were concatenated and inputted into other two parallel UNet models for segmentation of the shadow artifacts and NPAs. Overall, more than one-third of the papers in the literature used the basic UNet model for segmentation of vasculature and other pathological structures in the OCTA \textit{en face} images. It indicates that the UNet model is more adaptable for diverse applications due to the simple architecture and low computational cost, even though its segmentation accuracy can still be improved by designing more complex feature extraction mechanisms.

To improve the segmentation accuracy at multiple length scales, several existing works designed model frameworks with multiple information propagation paths as well as cross-modality feature extraction and fusion strategies \cite{89,90,93,94,95,96,97,97-1,102,102-1,102-2}. The dual-path feature processing model became a popular architecture, where one path focuses on the large-scale features of arteries and veins, and another path refines the continuous details of small vessels \cite{90,94,95,97}. For example, the OCTA-Net model proposed in~\cite{90} is a well-recognized model architecture for segmentation of the multi-scale vasculature. The ROSE dataset provided in this study has been a widely-used public dataset, as it contains not only \textit{en face} OCTA images, but also volumes of OCT and OCTA B-scan slices. As shown in Fig. 8a, the OCTA-Net model has a coarse stage for vessel segmentation and a fine stage for recover continuous details of small vessels. In the coarse stage, the backbone adopts a UNet model which uses ResNeSt blocks \cite{91} to replace the traditional convolution blocks. Two parallel information propagation branches are designed, where the deep UNet branch extracts the low-level pixel-wise features, and the shallow one aims to grade vessels in regions with poor contrast, more complex topological structures and relatively smaller diameters. Then, in the fine stage, the segmented pixel-wise and centerline-wise vessels and the raw OCTA image are combined to further improve the local vessel details using a split-based refined segmentation module~\cite{92}.

Other studies \cite{89,89-1,93,96} also used a UNet architecture as the backbone, and designed the individual blocks or the skipping connections by adding paths for extra features. For example, the study of \cite{96} applied three encoders with shared weights to extract multi-scale features from the OCTA \textit{en face} images acquired at the inner, superficial and deep retinal layers. Then, the gating modules, specifically designed for segmentation of vasculature, vascular junctions and FAZ, took the concatenation of the input images and the extracted features from different layers of the encoder for feature selection and fusion. Finally, the prediction heads were designed as two branches by combining the heatmap regression and grid classification.

To improve the global perception and reduce the redundant features generated by the convolution layers, different types of attention mechanisms were applied to UNet blocks \cite{90,98,99,100,101,102,103}. For example, the ResNeSt block in the OCTA-Net model \cite{90} used $k$ paths to extract spatial features. In each path, the input data $\mathbf{x}$ was splitted into $r$ patches, where each path was processed by $1\times 1$ and $3 \times 3$ convolutions. The features from $r$ patches were concatenated, and then channel attention was applied to identify the important features. Finally, features from $k$ branches were fused via weighted summation after being processed by the ResNeSt block (Fig. 8b (1)) \cite{91}. In~\cite{100}, the dynamic token aggregation attention (DTAA) block was designed as a modified MHSA model by adding a dynamic token aggregation (DTA) process before the calculation of attention values. Fig. 8b (2) shows the structure of one attention head, where the DTA process optimizes the token embedding matrix $W_{\theta}$, $W_{\phi}$ and $g(\mathbf{x})$, such that a large amount of redundant information in the tokens can be eliminated. Furthermore, the adaptive gated axial attention (AGAA) proposed in~\cite{102,104} decomposed self-attention into two self-attention modules along the height axis and the width axis. Compared with canonical self-attention in Fig. 3a (1), $R_{\theta}$, $R_{\phi}$ and $R_g$ in Fig. 8b (3) are designed for the axis-wise attention. Therein, $G_{\theta}$, $G_{\phi}$, $G_{g1}$ and $G_{g2}$ create gating mechanisms which filtered out less inaccurate positional embeddings.

\subsection{Weakly-supervised and Self-supervised Training Strategies}
\label{sec-4-2}
To alleviate the tedious work of labeling the retinal vessels, especially for the micro-vasculautre, previous studies proposed various weakly-supervised \cite{105,107,108,109} and self-supervised \cite{106,110,125} strategies and designed the corresponding network architectures. 

\begin{figure*}[!t]
\centerline{\includegraphics[width=0.7\columnwidth]{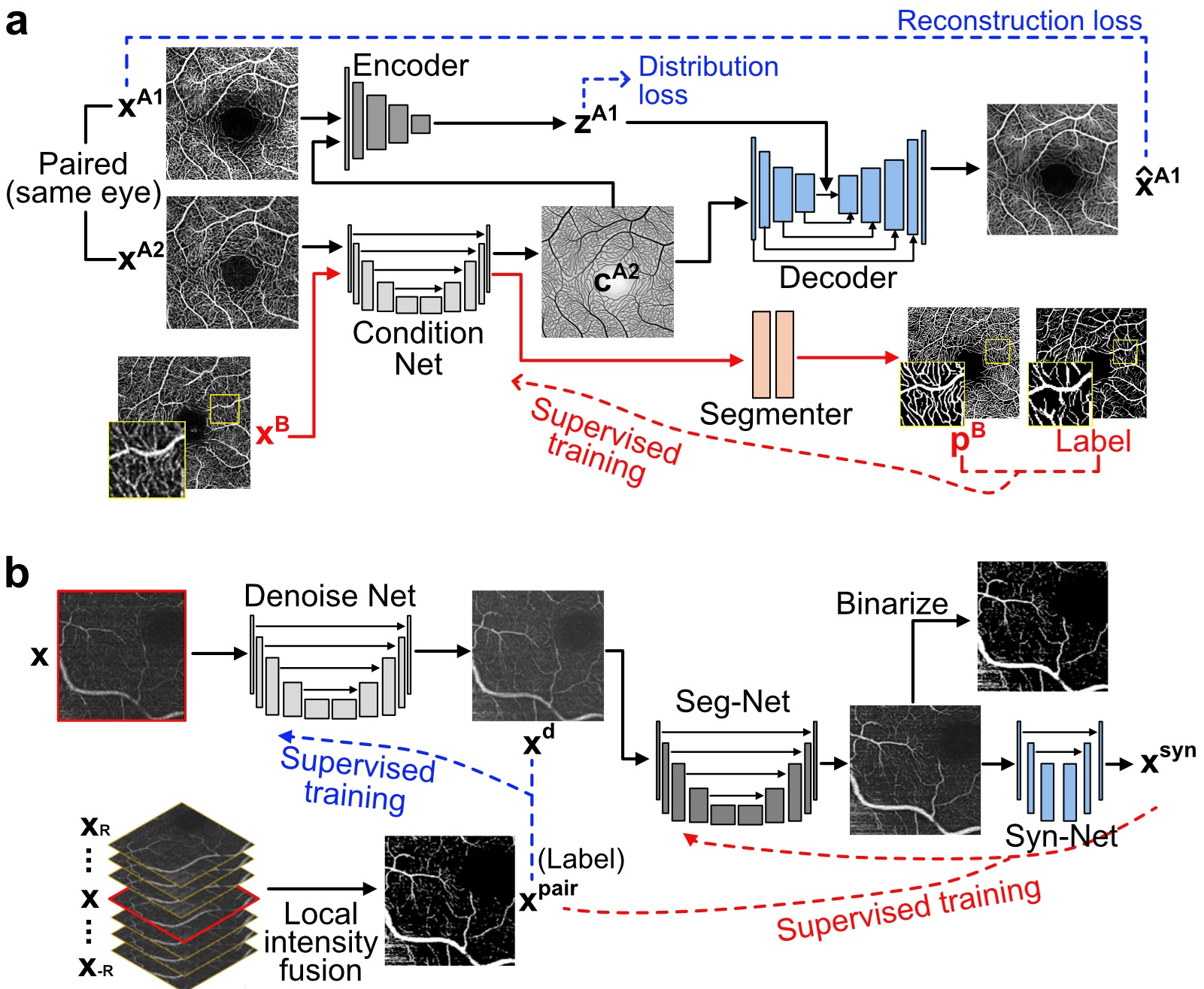}}
\caption{Weakly-supervised and self-supervised DL models for segmentation. \textbf{(a)} The weakly-supervised model uses a small amount of labeled OCTA data to train the condition network and segmenter, and uses a lot of paired OCTA data to extract the anatomical vascular features in the framework of CVAE \cite{107}. \textbf{(b)} The self-supervised model uses the registered image from local intensity fusion as the label to train the denoising network, segmentation network (Seg-Net) and image synthetic network (Syn-Net) \cite{106}.}
\label{fig9}
\end{figure*}

In the weakly-supervised models, some patches of label on the meaningful and informative areas or the scribble annotations were used. For example, in~\cite{107}, a conditional variational autoencoder (CVAE)-type of architecture was designed for weakly-supervised training, by using a small amount of training data with manual annotations and a large amount of paired imaged data without labels. As shown in Fig. 9a, first, a condition network (CN) and a segmenter are trained using the samples with labels. The segmenter only has two convolutional layers, and the conditional network captures the majority information about vascular shape. Next, paired image data $\mathbf{x}^{A1}$ and $\mathbf{x}^{A2}$, which are acquired at the same position of an eye using different devices or protocols, are input into the encoder and the CN, respectively. The latent feature image $\mathbf{c}^{A2}$ extracted by the CN contains information about vessels and capillaries, while the latent $\mathbf{z}^{A1}$ extracted by the encoder mainly focuses on the local contrast, artifacts and noises in $\mathbf{x}^{A1}$. $\mathbf{z}^{A1}$ and $\mathbf{c}^{A2}$ together reconstructed the artificial image $\hat{\mathbf{x}}^{A1}$ by the decoder. During training, only the reconstruction loss between $\mathbf{x}^{A1}$ and $\hat{\mathbf{x}}^{A1}$ and the KL divergence between the distribution of $\mathbf{z}^{A1}$ and the prior multivariate Gaussian distributions need to be minimized. Thus, the model is trained in a weakly-supervised manner without a large amount of annotations.

However, in practice, the idea of using a CVAE to extract common anatomical structures in two noisy OCTA images acquired by different devices could be unrealistic due to the limited data availability \cite{107}. Thus, in~\cite{106}, local intensity fusion (LIF), which is a kind of image registration technique, was applied to generate the paired image $\mathbf{x}^{pair}$ by using R-neighborhood images ${\mathbf{x}_{-R}, ..., \mathbf{x}_{R}}$. As shown in Fig. 9b, $\mathbf{x}^{pair}$ was then used for supervised training of the denoising network and the CVAE-type networks which consisted of segmentation network (Seg-Net) and image synthetic network (Syn-Net). Furthermore, another  study \cite{110,125} also proposed a self-supervised DL model based on the angiogenesis physics. The synthetic artificial OCTA images with labels of vessels were generated using a fast angiogenesis simulation and the CycleGAN-based style transfer learning. Then, a UNet model was trained using the synthetic images and labels for segmentation. The performance of the trained UNet was demonstrated using both the synthetic and real OCTA data. 

\subsection{Applications}
The vascular structures in the retina and choroid are critical indicators of many types of fundus diseases, such as diabetic retinopathy (DR) and age-related macular degeneration (AMD). Though a lot of DL models have been proposed to segment vasculature, their practical applications in medical research and clinics are still in the preliminary stage. Several previous works focused on quantifying the morphological differences between healthy and disease vasculature, such as the vessel diameter, density, length, tortuosity and the number of bifurcating nodes \cite{87,96,111,112,124}. For example, the study of \cite{112} observed a significant increase of capillary dilation in the retinal deep plexus in DR eyes. The study \cite{114} calibrated the distributions and densities of the flow voids (FVs) in the choroid, whose abnormal patterns indicated the pathological progression of central serous chorioretinopathy. The work in \cite{102-2} applied artificial perturbations on the volumetric OCTA data to study the plexus-wise vascular significance. They found that deep capillary plexus and choroid capillary plexus contained the most important morphological information about the AMD disease, such as the reduced capillary density and perfusion capability. It is also known that many pathological features of the arteriolar and venous vessels can be very different. For example, the capillary closure occurred largely on the arteriolar side, while morphological alternations of venous vessels such as looping were more prominent. Thus, besides segmentation of vasculature, classification of arterial-venous areas became the main goal of several studies \cite{83,83-2,83-3,113} to improve the diagnostic power. 

The appearance of non-perfusion areas (NPAs) and neovascularization (NV) are distinguishable signs in the later disease stages. Due to the damage of vascular endothelium by hyperglycemia, microaneurysms and dot intraretinal hemorrhage are generated at the early stage. As the disease progress, vasoconstriction and vascular occlusion lead to the formation of NPAs. In the last stage, the severe hypoxia causes an overexpression of vascular endothelial growth factor (VEGF), which ultimately lead to the NV in the retinal and choroid. Thus, a few previous works applied DL models for detection, segmentation and characterization of the NPAs \cite{55,85} and NV areas \cite{82,82-1,97-1,88}. For example, the study of \cite{97-1} tested the performance of a multi-path DL model on detection of the choroidal NV areas by using 10,566 OCT and OCTA images of 3135 eyes. Results showed that the overall performance of the model was reliable for clinical applications, though some variations under different diseases did exist.

Retinal vascular pathologies, such as the disrupted blood-retinal barrier, influence interstitial flow management, which causes local fluid accumulation and leads to the formation of macular edema (ME). In DR and AMD eyes, ME is one of the major causes of visual loss. Structural OCT has been widely used in clinics for diagnosis of ME based on the retinal thickness map and the volumetric OCT B-scan data. A recent study \cite{115} showed that the segmentation accuracy of ME and other intra-retinal an sub-retinal fluids can be improved by combing the OCT and OCTA information. 3D reconstruction and visualization of the fluid volumes was further achieved by registering all the images using the Bruch's membrane and large retinal vessels as the reference \cite{115}. 

Bruch's membrane is a $2-4 \mu m$ thick acellular matrix between the retinal pigment epithelium (RPE) and the choriocapillaries, which is a critical structural and functional support to the RPE. Many fundus disease such as AMD are related to the dysfunction of the Bruch's membrane caused by factors such as extracellular matrix degeneration and angiogenesis \cite{116}. A recent work \cite{117} applied a basic UNet model for BM segmentation using both OCT and OCTA B-scan images, by combining the anatomical knowledge about the intensity distributions and gradients of the retinal layers in these images. 

In clinics, many other important anatomical and pathological structures are used as reference for diagnosis and treatment or fundamental research. For example, the shape of the optical dis boundaries and the vascular morphology surrounding the optic nerve head are critical for evaluating disease progression like Glaucoma \cite{100-1,100-2}. Perfused and nonperfused microaneurysms are strongly associated with the extracellular fluid accumulation status, which are an early pathological indicator of the development of DR \cite{100-3}. The continuous development of OCTA devices and DL techniques will greatly accelerate developments in retinal disease clinical care and foundational ophthalmic science.

\section{Volumetric Reconstruction}
3D rendering of OCTA data provides intuitive and straightforward visualization results for clinicians to identify the shape and position of pathological features. It also helps reduce the interpretation inconsistency among different physicians. For example, 3D reconstruction of the retinal fluid volume enables accurately measuring the edema size and localized the neovasculature position, which provides useful reference for identifying the severity of the disease and guiding the high-precision vitreo-retinal surgery. Furthermore, the high-quality structural model of the 3D vasculature can be used in hemodynamics analysis and neurovascular coupling analysis, illustrating how oxygen and nutrients are delivered to the optic nerve cells in the fundus tissues.

\begin{figure*}[!t]
\centerline{\includegraphics[width=0.8\columnwidth]{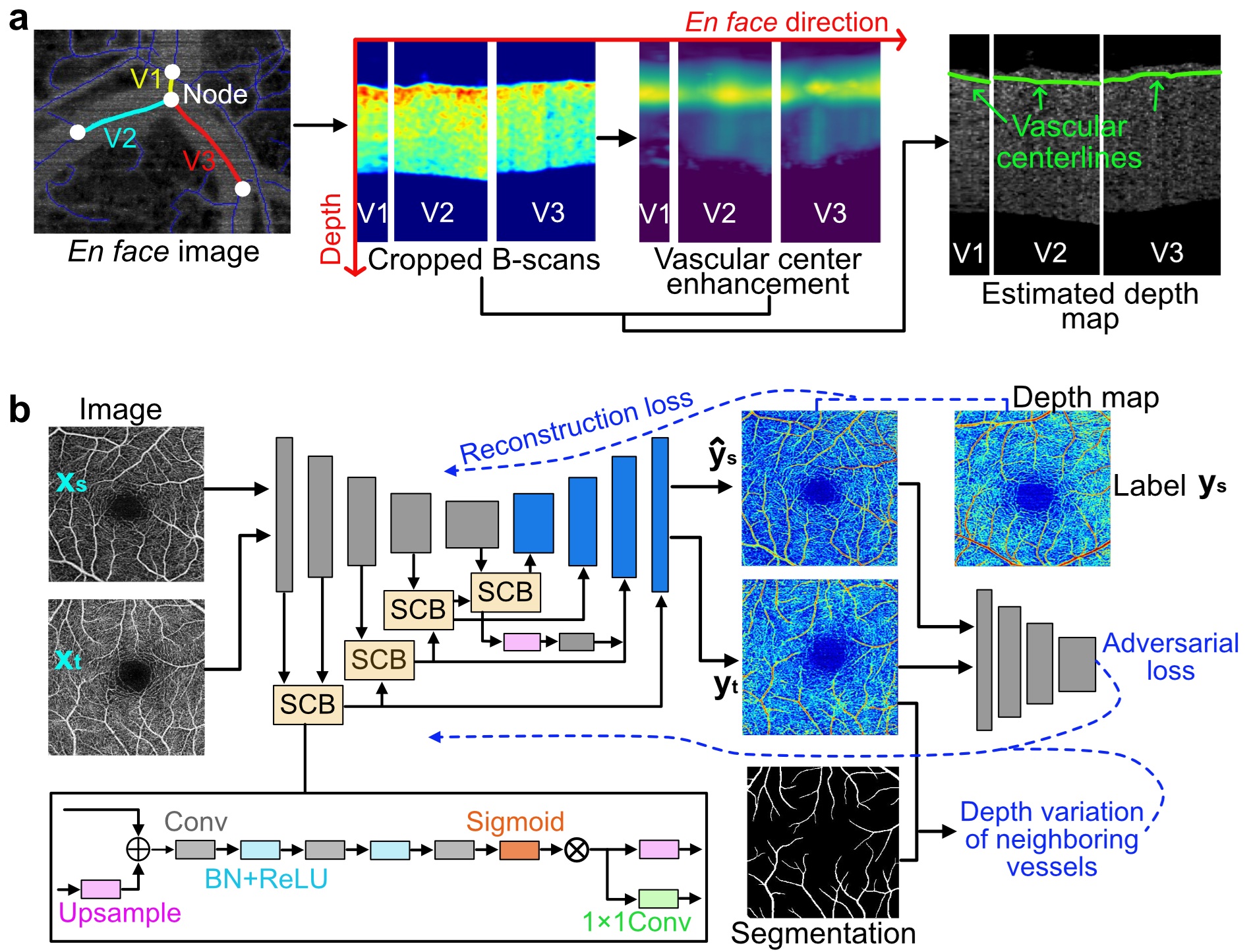}}
\caption{Quantification of the depth map of vessels. \textbf{(a)} The depth map of each vascular segment calculated based on the intensity distributions in the B-scan slices \cite{120}. \textbf{(b)} A style-transfer DL model for prediction of the depth map based on the \textit{en face} OCTA images acquired by different devices \cite{122}.}
\label{fig10}
\end{figure*}

3D geometry data comprises three primary types: point clouds, surface meshes and voxel grids \cite{118}. Point clouds are data points measured in the 3D space, which contained the intensity information. The denser the points, the more detailed small-scale features can be represented by the point clouds. Point clouds data are memory efficient for storage. However, they are irregular and unstructured data, where the connectivity among data points is lost and the data is permutation invariant. Comparatively, surface meshes and voxel grids are regular graph data, which contain richer information such as the spatial constraints among neighboring data points and the shape of the whole object. For example, several aforementioned studies have shown the 3D shape of the edema and vasculature using a stack of denoised or segmented \textit{en face} OCT and OCTA image slices \cite{1,44,52,89,123,124}.  Furthermore, analysis of blood flows and oxygen distributions by using computational models relies on the regular graph data.

Point clouds of OCTA data can be converted to regular graph data by compensating the lost information, including the depth information and spatial connectivity among data points. We review four recent studies to exemplify the problem of 3D shape reconstruction of the retinal vasculature \cite{119,120,121,122}. Two studies were based on the graph analysis and mesh reconstruction techniques \cite{119,120}, and the other two used DL models to estimate the depth map of vessels~\cite{121,122}.  With non-DL-based techniques, the study of \cite{119} developed a cuboid intersection method for reconstruction of triangular meshes of vascular surfaces from the point clouds data. Fig. 10a shows a novel method proposed by~\cite{120} for extraction of the depth map using the volume of OCTA B-scan slices. This kind of signal processing algorithms were implemented in the AngioPlex\textsuperscript{\textcircled{R}} OCT Angiography software, such that a vascular depth map can be generated automatically with the OCTA B-scans and \textit{en face} images after the experimental measurement. These depth maps were treated as key training labels for the DL models in studies of \cite{121,122}, which designed DL models (Fig. 10b) for predicting the depth maps based on \textit{en face} images measured by the same or different devices. 

Specifically, to estimate the depth map, the vascular centerlines and the bifurcation nodes were firstly extracted by hands or skeletonization algorithms using \textit{en face} OCT and OCTA images. For example, in Fig. 10a, three vessel components between two nodes (V1, V2 and V3) are shown. The OCTA B-scan images corresponding to the three vessel components are selected, where the red color represents the maximum signal intensity. By considering the shadow effect of vessels on the beneath signals, the enhancement image of the vascular center is calculated. Based on the B-scan images and the center-enhancement images, the centerlines of vessels in both \textit{en face} and depth directions can be accurately estimated, and the depth map is generated. Assuming that vessels have tubular shape with the same diameter, the 3D vasculature can be reconstructed \cite{120}. The study \cite{122} designed a domain-transfer DL model, which  predicts the depth map of the vasculature using only the \textit{en face} OCTA images acquired by different devices. The model adopts a UNet as the backbone model, and incorporates SCB blocks, whose detailed structure is shown in the black box, for providing refined structural information of the decoder. Input images $\mathbf{x}_s$ and $\mathbf{x}_t$ are source images and target images, respectively. The source images carry depth maps $\mathbf{y}_s$ generated by the OCTA device as the ground truths. Transferring the model prediction $\hat{\mathbf{y}}_t$ to the target domain is achieved via adversarial training and continuity correction by minimizing the depth variation in a sliding window.  

So far, there have been only few preliminary works about developing 3D structural models for the vasculature and investigating the mass circulation dynamic in the retinal tissues. Currently, generative DL models have been widely applied for 3D reconstruction of human organs, such as bone, brain and tumor,  using CT and MRI images for precision diagnosis, surgery or regenerative treatments \cite{118,126,127}. Though the structural segmentation and reconstruction of the multi-scale vascular networks in the fundus is certainly more complex and largely influenced by the spatial resolution of the measurement, this research direction would attract increasing attention due to its practical importance.  

\section{Public OCTA dataset}
A large amount of real-world OCTA data with good image quality and accurate delineations can significantly improve the performance of DL models. However, compared with other image types such as fundus photographs, the number of publicly available OCTA datasets remains limited. The quality of OCTA images vary over different dataset, since it is influenced by the acquisition devices and protocols. We list seven major public OCTA datasets~\cite{81,9284503,128,129,130,131,132} in Table~\ref{tab1}.

Among the most widely-used public datasets for retinal OCTA analysis are ROSE~\cite{9284503} and OCTA-500~\cite{130}. The ROSE (Retinal OCTA SEgmentation) dataset, released by the Chinese Academy of Sciences, comprises 229 OCTA images with manual vessel annotations at either centerline or pixel level. It consists of two subsets (ROSE-1 and ROSE-2) acquired using different devices (Table~\ref{tab1}, ROSE). Both subsets provide $3 \times 3 mm^2$ fovea-centered scans for high resolution. ROSE-1 includes centerline-level and pixel-level labels for angiograms of the superficial vascular complex (SVC), deep vascular complex (DVC), and the combined SVC and DVC from 39 samples, with images sized $304\times 304$ pixels. ROSE-2 contains 112 SVC angiograms from 112 eyes as $840\times 840$ pixel grayscale images. In contrast, OCTA-500 from Nanjing University of Science and Technology offers a larger dataset comprising OCTA images from 500 subjects acquired using a single device. This dataset provides images and annotations in two fields of view ($6\times6$ mm$^2$ and $3\times3$ mm$^2$), featuring both 3D OCT volumes and OCTA B-scans (Table 1). OCTA-500 offers comprehensive annotations including four text-based labels and seven types of 2D/3D segmentation labels. Among the 7 types of segmentation labels, labels of capillary, artery, vein and 3D FAZ enable integrated analysis of capillary, arterial, venous, and FAZ segmentation within a unified framework~\cite{130}.

The largest public OCTA dataset to date was released by Zhongshan Ophthalmic Center at Sun Yat-sen University~\cite{132}.  This dataset contains $25,665$ \textit{en face} OCTA images of retinal vascular acquired with a single device and uniformly resized to $320 \times 320$ pixels. It comprises four subsets categorized by field of view, namely, Subset sOCTA-$3\times3$-10k containing $10,480$ $3\times3 mm^2$ SVC layer images, Subset sOCTA-$6\times6$-14k containing $14,042$ $6\times6 mm^2$ SVC layer images, Subset sOCTA-$3\times3$-1.1k-seg containing $1,103$ $3\times3 mm^2$ DVC images, and Subset sOCTA-$6\times6$-1.1k-seg containing $1,143$ $6\times6 mm^2$ DVC images. Additionally, an external test set from West China Hospital is included. The dataset provides image-level classification labels and pixel-level FAZ segmentation masks.

Compared to the three major datasets, other public datasets (see Table~\ref{tab1}) are either relatively small in terms of dataset scale (e.g., 55 \textit{en face} images~\cite{81} altogether or 36 OCTA images per subset~\cite{128}), or restricted in resolution and annotations (e.g., only offering $6\times6 mm^2$ SVC image~\cite{129}). Such limitations may constrain the development and validation of complex DL models for clinical applications using these datasets.

Besides the public OCTA datasets listed in Table~\ref{tab1}, examples of other less-utilized open datasets include the Diabetic Retinopathy Analysis Challenge (DRAC)~\cite{DRAC} and the Retinal OCTA Diabetic Retinopathy dataset (ROAD)~\cite{ROAD-ds}. DRAC~\cite{DRAC} comprises 1,103 ultra-wide OCTA (UW-OCTA) images ($1,024\times 1,024$ pixels) captured with a single device. It supports three tasks: image quality assessment, DR lesion segmentation, and DR grading. For the DR grading task, ophthalmologists provides text labels for images as non-DR, NPDR, or PDR, with 611 images for training. The segmentation task includes a relatively small number of 109 training and 65 test samples with ground truth masks. In comparison, ROAD~\cite{ROAD-ds} focuses on segmentation and grading, offering a larger dataset of 1,200 non-DR images, 1,440 DR images and 1,440 segmentation masks. It also provides text labels for three grades including non-DR, mild non-proliferative diabetic retinopathy (NPDR), and proliferative diabetic retinopathy (PDR). In addition, a multimodal dataset was released by~\cite{bidwai2024multimodal} to include OCTA images and 111 color fundus images. However, similar to its uni-modal counterpart~\cite{bidwai2024singlemodal} (containing 268 OCTA-only images with labels), only text labels regarding the disease categories were provided,  limiting the dataset's potential applications.

Beyond the public datasets summarized in Table~\ref{tab1}, several noteworthy private OCTA datasets also deserve attention. For instance, a study developed a deep learning system for neovascularization of the optic disc (NVD) identification using two large private datasets~\cite{157}, where one contains 24,576 OCTA images from 96 NVD patients, and another comprises 15,360 OCTA images from 60 NVD patients. All images were acquired via $6\times 6$ $mm^2$ optic nerve-centered volume scans, with manual annotations of NVD regions. Although these datasets remain unavailable due to ongoing research, they are reportedly accessible upon request from the authors of~\cite{157}.

\newcolumntype{L}{>{\RaggedRight\hangafter=1\hangindent=0em}X}	
\begin{sidewaystable}[t]
\small
\centering
\label{table-A1}
\renewcommand{\tablename}{Table}
\caption{Public OCTA dataset.}
\hspace{1pt}
\setlength{\tabcolsep} {3pt}

\begin{tabularx}{\textwidth}{lp{5cm}p{11cm}L }		
\toprule[1pt]
\textbf{Dataset} & \hspace{1.5cm}\textbf{Link} & \hspace{4cm}\textbf{Data info}  \\
\midrule[0.5pt]
Giarratano \cite{81}    & \url{https://datashare.ed.ac.uk/handle/10283/3528}  &  55 \textit{en face} images of 11 healthy eyes, at five regions of interest: superior, nasal, foveal, inferior, and temporal. Acquired by RTVue-XR Avanti OCT system (OptoVue), $304 \times 304$ repeated A-scans. \vspace{0.5cm}\\

ROSE \cite{9284503} &  \url{https://imed.nimte.ac.cn/dataofrose.html} & \textbf{(1) ROSE-1:} 117 \textit{en face} images of 13 healthy and 26 Alzheimer’s disease subjects, $3 \times 3 mm^2$ images of superficial (SVC), deep (DVC) and inner vascular complexes (IVC), centered at the FAZ. Acquired by  RTVue XR Avanti SD-OCT system (Optovue) equipped with AngioVue software, $304 \times 304$ repeated A-scans. \textbf{(2) ROSE-2:} 112 \textit{en face} images of 112 disease eyes, centered at the FAZ, $3 \times 3-mm^2$ SVC images. Acquired by Heidelberg OCT2 system with Spectralis software (Heidelberg Engineering), $512 \times 512$ repeated A-scans. \vspace{0.5cm}\\

OCTAGAN \cite{128}    & \url{http://www.varpa.es/research/ophtalmology.html#cloud} & 212 \textit{en face} images of 36 healthy and 17 DR eyes, $3 \times 3-mm^2$ SVC and DVC images. $6 \times 6-mm^2$ SVC and DVC images, centered at the FAZ. Acquired by DRI OCT Triton (Topcon), $320 \times 320$ repeated A-scans. \vspace{0.5cm}\\

FAZID \cite{129} & \url{https://www.openicpsr.org/openicpsr/project/117543/version/V2/view}  &304 \textit{en face} images of 88 healthy, 107 DR eyes and 109 myopic eyes. $6 \times 6-mm^2$ images centered at the FAZ. Acquired by Cirrus 5000 Angioplex (Carl Zeiss Meditec Inc.), $420 \times 420$ repeated A-scans. \vspace{0.5cm}\\

OCTA-500 \cite{130}  & \url{https://ieee-dataport.org/open-access/octa-500}  &\textbf{(1) OCTA-6mm: } 300 OCT and OCTA volumes, $6 \times 6 \times 2-mm^3$, 209 are disease eyes (DR, AMD, CNV, CSC, RVO) and 91 are healthy eyes. \textbf{(2) OCTA-3mm:} 200 OCT and OCTA volumes, $3 \times 3 \times 2-mm^3$, 40 are disease eyes (DR, AMD, CNV) and 160 are healthy eyes. Acquired by RTVue-XR Avanti OCT system (OptoVue). OCTA-6mm used $400 \times 400 \times 640$ repeated A-scans. OCTA-3mm used $304 \times 304 \times 640$ repeated A-scans.\\
\bottomrule[1pt]
\end{tabularx}
\label{tab1}
\end{sidewaystable}

\begin{sidewaystable}[thp]
\small
\centering
\renewcommand{\tablename}{Table 1}
\hspace{1pt}
\setlength{\tabcolsep} {3pt}

\begin{tabularx}{\textwidth}{lp{5cm}p{11cm}L }		
\toprule[1pt]
\textbf{Dataset} & \hspace{1.5cm}\textbf{Link} & \hspace{4cm}\textbf{Data info}  \\
\midrule[0.5pt]

SOUL \cite{131} & \url{https://doi.org/10.6084/m9.figshare.24893358.v3}  & 53 \textit{en face} images of Branch Retinal Vein Occlusion (BRVO) eyes, $6 \times 6-mm^2$ SVC layer. Acquired by Optovue Angio OCT RTVueXR. Acquisition protocols are not specified.\vspace{0.5cm}\\

COIPS \cite{132} & \url{https://doi.org/10.5281/zenodo.5111975} and \url{https://doi.org/10.5281/zenodo.5111972}  & \textbf{(1) sOCTA-3$\times$3-10k:} 10480 \textit{en face} images, $3 \times 3 mm^2$ SVC layer. \textbf{(2) sOCTA-6$\times$6-14k:} 14042 \textit{en face} images, $6 \times 6 mm^2$ SVC layer. \textbf{(3) sOCTA-3$\times$3-1.1k-seg:} 1101 \textit{en face} images, $3 \times 3 mm^2$ SVC layer. \textbf{(4) dOCTA-6$\times$6-1.1k-seg:} 1143 \textit{en face} images, $6 \times 6 mm^2$ DVC layer. Acquired by DRI OCT Triton (Topcon), $320 \times 320$ repeated A-scans.\\
\bottomrule[1pt]
\end{tabularx}
\end{sidewaystable}

\clearpage
\section{Discussion and Open Issues}
In summary, among the diverse DL models for denoising, segmentation, and volumetric rendering of OCTA images, those developed prior to 2021 represent an early developmental phase, predominantly employing basic architectures such as CNNs, U-Nets and various attention mechanisms. Although these models outperformed traditional methods (e.g., thresholding-based algorithms), their reliance on labor-intensive expert annotations limited real-world applicability, due to the requirement for large-scale annotated datasets to train more complex architectures effectively. More recent studies have introduced multimodal fusion frameworks that integrate multiple feature processing paths—each containing basic DL models (e.g., CNN or U-Net) to extract structure-specific features hierarchically (see our discussion in Section~\ref{sec-4-1}). These frameworks enable fusion of multi-source data (e.g., OCT/OCTA images, reflectance maps, and retinal thickness maps) at different processing stages, enhancing the inference strength of DL models.

Furthermore, recent studies have also incorporated OCTA-specific characteristics, such as vascular structures and consecutive B-scan features, to develop domain-specific DL models. For instance, spatial-temporal B-scan sequences encode prior physical knowledge (e.g., vasculature continuity), which enables model architectures leveraging these attributes to improve performance while reducing annotation burdens (see our discussion in Section~\ref{sec-4-2}). Preliminary integrations of angiogenesis dynamics and light physics also demonstrate that such first-principle mechanisms provide valuable priors for both supervised and unsupervised DL models, despite being in early development.

However, despite these advances, OCTA image analysis still faces critical challenges. Firstly, supervised DL models using basic DL component designs (notably U-Net variants) remain the most reliable approach for processing disease images with significant variations in patterns and retinal structures, primarily because vasculature segmentation accuracy heavily depends on annotation quality. Unlike general-purpose vision tasks (e.g., ImageNet), the lack of large-scale datasets hinders the adoption of larger models like Vision Transformers~\cite{Salmanvit} and Vision Mambas~\cite{liu2024visionmambacomprehensivesurvey}.  In contrast, fundus photography benefits from abundant public datasets, enabling extensive DL development for ocular disease diagnosis~\cite{LI2021101971,10596287, li2024predicting}. This disparity highlights the potential of cross-modal transfer learning, where models trained on fundus images could generate pseudo-labels for self-supervised OCTA analysis, as demonstrated in multi-modal signal processing research such as~\cite{Zhao_2018_CVPR}. 

Secondly, the prevalence of compact/lightweight DL models underscores the need for OCTA-specialized foundation models capable of integrating vision-language understanding. For general-purpose tasks, contrastive language-image pre-training (CLIP)~\cite{clip} excels in image understanding, especially in the tasks of out-of-distribution context understanding~\cite{10655065, s23218757}. For medical images, integrated frameworks like DeepDR-LLM~\cite{li2024integrated} demonstrate multimodal diagnostic potential by combining LLMs with retinal imaging through training based on multi-country datasets comprising more than one million retinal fundus photographs. Such advancement also inspires the potential of integrating OCTA images as one informative modality into the languarage-image DL framework. OCTA’s unique value lies in its ability to visualize vascular structures from superficial to deep retinal layers as well as the choroidal layer, including anterior segment angiography of high-flow regions. Integrating OCTA with ophthalmological knowledge and biomechanics could elucidate disease mechanisms and accelerate therapeutic development.

Thirdly, most available OCTA datasets and DL models face cross-device generalization challenges due to significant image quality variations caused by the differences in OCTA instruments (e.g., Spectralis vs. AngioVue) and scanning protocols~\cite{PROTSYK2022421, tsai2024comparison}. Currently, most DL models are typically trained on single-scanner data, and thus struggle with generalization to other devices because of divergent artifact patterns (e.g., motion artifacts vs. low SSI) and signal characteristics (e.g., variations in B-scan rates). These limitations confine applications of OCTA-based DL model design to proof-of-concept stages. Addressing this issue requires unsupervised image-to-image translation techniques, particularly those leveraging the power of generative models such as CycleGANs and diffusion models~\cite{nouri2023addressing}, which enable device-to-device adaptation without paired training data. Concurrently, establishing standardized cross-device datasets is critical to developing targeted DL-based solutions for heterogeneous data sources.

\section{Conclusion}
In this survey, we have provided a comprehensive analysis of deep learning (DL) model-based applications in OCTA image processing over the past five years. Our review has been focused on three critical tasks: denoising, segmentation and volumetric rendering. We have chronologically reviewed prevalent DL architectures built upon foundational building blocks including CNNs, U-Nets and various attention models. We have also highlighted persistent challenges, including dataset scarcity (especially compared to fundus photography), computational demands for 3D processing, and limited generalizability across OCTA devices due to domain shifts in image quality and artifacts. We have also provided insights into technological directions for future advancements, including the potential integration of optical physics principles (e.g., light attenuation models) and advanced vision/language-vision models to enhance robustness and clinical applicability.

\bibliographystyle{elsarticle-num-names} 
\nocite{*}
\bibliography{elsarticle-bib}





\end{document}